\documentstyle[aas2pp4]{article}
\newcommand{\cgro}{{\it CGRO}}
\newcommand{\approxlt}{\mbox{$\;^{<}\hspace{-0.24cm}_{\sim}\;$}}

\begin{document}
\title{A Search for Non-triggered Gamma Ray Bursts in the BATSE Data Base}

\author{Jefferson M.\ Kommers,\altaffilmark{1} Walter H.\ G.\
  Lewin,\altaffilmark{1} Chryssa Kouveliotou,\altaffilmark{2,3} Jan van
  Paradijs,\altaffilmark{4,5} Geoffrey N.\ Pendleton,\altaffilmark{4}
  Charles A.\ Meegan,\altaffilmark{3} and Gerald J.\ Fishman\altaffilmark{3}}

\altaffiltext{1}{Department of Physics and Center for Space Research,
  Massachusetts Institute of Technology, Cambridge, MA 02139}
\altaffiltext{2}{Universities Space Research Association, Huntsville, AL 35800}
\altaffiltext{3}{NASA/Marshall Space Flight Center, Huntsville, AL
  35812}
\altaffiltext{4}{University of Alabama in Huntsville, Huntsville, AL
  35812}
\altaffiltext{5}{University of Amsterdam, Amsterdam, Netherlands}

\lefthead{DRAFT}

\begin{abstract}
  We describe a search of archival data from the Burst and Transient
  Source Experiment (BATSE).  The purpose of the search is to find
  astronomically interesting transients that did {\it not} activate
  the burst detection (or ``trigger'') system onboard the spacecraft.
  Our search is sensitive to events with peak fluxes (on the 1.024 s
  time scale) that are lower by a factor of $\sim$ 2 than can be
  detected with the onboard burst trigger.  In a search of 345 days of
  archival data, we detected 91 events in the 50--300 keV range that
  resemble classical gamma ray bursts but that did not activate the
  onboard burst trigger.  We also detected 110 low-energy (25--50 keV)
  events of unknown origin which may include activity from SGR
  1806$-$20 and bursts and flares from X-ray binaries.  This paper
  gives the occurrence times, estimated source directions, durations,
  peak fluxes, and fluences for the 91 gamma ray burst candidates.
  The direction and intensity distributions of these bursts imply that
  the biases inherent in the onboard trigger mechanism have not
  significantly affected the completeness of the published BATSE gamma
  ray burst catalogs.
\end{abstract}


\section{Introduction}
\setcounter{footnote}{0}
\label{sec:intro}
Since 1991 April 19 the Burst and Transient Source Experiment
(BATSE) on the {\it Compton Gamma Ray Observatory} (\cgro) has been
detecting gamma ray bursts (GRBs) and other high-energy transients
with unprecedented sensitivity (\cite{Fishman89}; \cite{Batse1b};
\cite{Batse3b}).  The 1122 GRBs in the 3B catalog show an isotropic
angular distribution and a spatially inhomogeneous intensity
distribution (\cite{MeeganNature92}; \cite{Batse3b}).  Despite
extensive analysis, however, the origin of GRBs remains unknown; see
recent reviews by Fishman \& Meegan (1995), Briggs (1995), and
Hartmann (1995).

The detection of GRBs and other high-energy transients with BATSE is
controlled by a real-time burst detection algorithm running onboard
the spacecraft (\cite{Fishman89}).  The onboard computer continuously
monitors the count rates in each of the eight Large Area Detectors
(LADs).  When the count rates exceed a certain threshold, the computer
signals a ``burst trigger'' and data are collected at high temporal
and spectral resolution for a limited time interval.  Even in the
absence of a burst trigger, however, data are recorded at lower
resolution in the continuous data types.  For most of the mission, the
criteria for a burst trigger have been that the 50--300 keV count
rates in two detectors simultaneously increase by more than 5.5 times
the expected root-mean-square background fluctuations on any of three
time scales: 64 ms, 256 ms, or 1024 ms.  The average background rate
for each detector is recomputed every 17.408 s (\cite{Fishman89}).

By definition all of the GRBs listed in the 1B, 2B, and 3B catalogs
satisfy the requirements for a burst trigger (\cite{Batse1b};
\cite{Batse3b}).  Other transient phenomena that are unrelated to GRBs
can also lead to a burst trigger.  Examples include solar flares,
terrestrial magnetospheric disturbances, bursts and flares from X-ray
binaries, and activity from soft gamma ray repeaters (SGRs).  Such
events are classified appropriately by the BATSE team.

A GRB or other transient phenomenon may have characteristics such that
it does {\it not} lead to a burst trigger onboard the spacecraft but
it nevertheless leaves a statistically significant signal in the
continuous data.  For example, a GRB or other transient may be too
faint to achieve the necessary statistical significance for a trigger;
it may have a time profile that biases the onboard background average;
or it may have too few counts in the 50--300 keV range.

A burst can also occur while the onboard trigger is disabled for
technical reasons.  Following a burst trigger, the high resolution
data collected during the burst accumulation interval are gradually
telemetered to the ground during the following 90 minutes.  During
this read out period the onboard burst trigger is disabled on the 256
ms and 1024 ms time scales, and the 64 ms threshold is set to the
maximum rate of the burst being read out.  The onboard burst trigger
is also disabled when the spacecraft passes through regions with a
high probability of triggering on atmospheric particle precipitation
events (\cite{Batse1b}).

In this paper, we describe a retrospective search of the archival
continuous data from BATSE for statistically significant GRBs and
other transients that did {\it not} cause a burst trigger onboard the
spacecraft.  A search for these ``non-triggered'' (or ``untriggered'')
events in the 50--300 keV range is expected to find GRBs that are
generally fainter than those cataloged previously.  A concurrent
search for non-triggered events in the lowest discriminator channel
(25--50 keV) is expected to find activity from other astronomical
sources, including bursts and flares from X-ray binaries and activity
from SGRs.  This ongoing project is an extension of previous work by
Rubin et al. (1993), Van Paradijs et al.  (1993), and Kommers et al.
(1996).  Other retrospective searches for GRBs in the BATSE data
(using techniques different from those described here) have been
discussed by Skelton \& Mahoney (1994) and by Young et al.  (1996).

\section{Search Algorithm}
\label{sec:searchalg}
A retrospective search of archival data can take advantage of burst
detection algorithms that would have been impractical to implement
onboard the spacecraft.  The choice of a detection scheme to look for
non-triggered events therefore involves trade-offs between the
detection efficiency of the method for a given class of events and the
resources (both computational and human) needed to implement it.  In
this section we describe a search algorithm which is loosely based on
the one used onboard the spacecraft but which has proved more
sensitive.  The next section will discuss its efficiency for detecting
transients with certain characteristics.

We will refer to all events detected by our off-line search of
archival data as ``laboratory triggers.''  The events previously
detected by the onboard burst trigger mechanism will be called ``onboard
triggers.''  Some onboard triggers will be flagged by our off-line
search and so they will also be laboratory triggers.  Events that were
detected {\it only} by our off-line search will be called
``non-triggered events.''

For most of the mission the onboard trigger criterion has required
that the count rate in two detectors simultaneously increase by at
least 5.5$\sigma_B$ above the nominal background level, where
$\sigma_B$ is the standard deviation of the expected background counts
due to counting statistics.  As a result, the BATSE detectors provide
anisotropic sky exposure over short time periods (\cite{Fishman89};
\cite{Batse1b}).  The cosine-like change in the detectors' effective
area with source viewing direction causes the onboard trigger to be
less sensitive to faint bursts with directions directly in front of
one of the detectors than to ones with directions mid-way between two
detector normals (\cite{brock91}).  For example, a $10 \sigma_B$ event
occurring directly in front of a detector may produce only a $3.5
\sigma_B$ signal in the second most brightly illuminated detector and
it would thus fail to trigger onboard.  On the other hand, the same
event incident along a direction mid-way between two detector normals
would register approximately $7.1 \sigma_B$ in both detectors and
would comfortably cause an onboard trigger.

The onboard trigger mechanism relies on a background rate that is
computed during a 17.408 s time interval occurring before the time bin
being tested.  A rising or falling background can therefore bias the
background estimate to be too low or too high, respectively.  A slowly
rising transient may itself bias the background estimate upwards to
such an extent that it fails to cause an onboard burst trigger, even
though it is otherwise intense enough to be above the minimum
detection threshold.

Our retrospective search procedure partially combats the directional
detection anisotropy and the rising/falling background bias.  We form
a time series to be searched by combining the relevant energy channels
from the DISCLA data.  This data type provides the count rates in each
detector integrated over 1024 ms time bins.  Four discriminator energy
channels (numbered 1 through 4) are available: 25--50 keV, 50--100
keV, 100--300 keV, and $>300$ keV.  When searching for GRBs, the sum
of channels 2 and 3 (50--300 keV) gives optimal sensitivity.  When
searching for low-energy transients such as bursts and flares from
X-ray binaries or SGRs the lowest energy channel (channel 1, 25--50
keV) is most sensitive.  A sum of channels 1,2, and 3 provides a
further ``catch-all'' search.  After summing the appropriate energy
channels, the time series are rebinned in time (if necessary) to
search on time scales longer than the 1024 ms DISCLA sampling period.
The resulting time series are then searched sequentially to see if any
data meet our laboratory trigger criteria.

To signal a laboratory trigger, we first determine from the time
series being searched a nominal background level for each detector.
To estimate $B_d(k)$, the number of background counts expected in
detector $d$ for the time bin $k$, we use a linear fit to the data in
time bins $k-N_b, \ldots, k-1$ and $k + 1 + N_b, \ldots, k + 2N_b$.
Here the number of background bins, $N_b$, is specified for each time
scale.  Unlike the onboard trigger, the off-line search uses data
before and after the time bin being tested.\footnote{The bin, $k$, being
  tested is not centered in the gap of $N_b$ bins which separates
  the fitted intervals.  This somewhat arbitrary choice evolved out of
  various triggering schemes which were tried, including one where the
  background was always estimated based on the $N_b$ bins immediately
  prior to the one being tested.  The scheme chosen performed well on
  14 days of data that were used to test various laboratory trigger
  criteria.} This method reduces the bias (discussed above) caused by
slowly rising or falling background levels.

Temporally contiguous data are not always available because of
telemetry gaps and spacecraft passages through regions of high
particle flux, such as the South Atlantic Anomaly (SAA).  In such
cases the background estimate discussed here cannot be formed.  We
only test bins for which we can estimate the background with the above
procedure, so our search is not sensitive to bursts that occur during
the $N_b$ bins after, or the $2N_b$ bins before, a data gap.

Let $C_d(k)$ be the measured number of counts in time bin $k$ for
detector $d$.  We define the ``significance'' of that detector to be
$S_d(k) = [C_d(k) - B_d(k)]/\sqrt{B_d(k)}$.  Our laboratory trigger
criteria are that the two greatest values of the 8 significances
$S_d(k)$, call them $s_1$ and $s_2$, must be such that $s_1 \geq s_2
\geq M$ and $s_1 + s_2 \geq \Sigma$.  These criteria ensure that at
least two detectors simultaneously experience a statistically
significant upward fluctuation, but they are also more sensitive to
events incident along detector normals than the onboard criteria.

The values chosen for $M$, $\Sigma$, and $N_b$ for the searches
reported here are shown in Table \ref{tbl:TriggerParameters}.  They
were chosen to keep the actual number of detections per day of data
searched (due to the activity of real sources) at a manageable level
of about 20 per day.

A more sensitive search could be conducted using other laboratory
trigger criteria.  For example, the effective detector area achieved
by adding the rates in each set of 3 contiguous detectors would
produce time series with higher signal-to-noise ratios for activity
from real sources.  The caveat to a more sensitive search is the
corresponding increase in the rate of ``false triggers'' due to solar
activity, variability from X-ray binaries, particle precipitation
events, Earth occultation steps, and phosphorescence spikes (which
occur when high-energy particles interact in the detectors).  During
outbursts of bright X-ray binaries, our laboratory trigger 
has detected hundreds of events per day due to variability from
Vela~X-1, A0535$+$26, and GRO~J0422$+$32.

For our laboratory trigger criteria, the number of detections expected
from statistical fluctuations alone can be estimated by considering
the ``phase space'' defined by $s_1$ and $s_2$.  The number of counts
in each time bin of the DISCLA data type is large enough that the
Poisson statistics can be treated in the Gaussian approximation, so
that the $S_d(k)$ are independent and normally distributed with zero
mean and unit variance.  The laboratory trigger criteria define a
region $A$ in the $(s_1,s_2)$ plane in which measured values $s^{*}_1$
and $s^{*}_2$ will be flagged by our search.  This area is sketched in
Figure \ref{fig:trigarea}.

The probability $P_{stat}$ that a randomly selected set of 8 significances
$S_d$ will meet or exceed the laboratory trigger criteria is
estimated by first integrating the bivariate normal distribution
(centered on the origin) over the allowed area $A$ in the $(s_1,s_2)$
plane.  Next, we multiply by 8 ways to select the most significant
detector ($s_1$) and the remaining 7 ways to select the second most
significant ($s_2$).  The final expression is
\begin{equation}
P_{stat} = \left[ \int\!\!\int_{A} \frac{1}{2\pi} \exp \left( - \frac{s_1^2 +
      s_2^2}{2} \right) \, ds_1 \, ds_2 \right] \times 7 \times 8.
\end{equation}
If there are $N$ time bins searched per day, the number of detections
expected from purely statistical fluctuations is $P_{stat} \times N$
per day.  This quantity is shown in the last column of Table
\ref{tbl:TriggerParameters} for a representative value of $N = 5.0
\times 10^{4}$.  Although the searches are not statistically
independent, we would expect no more than 8 detections due solely to
statistical fluctuations in a search of 100 days of data.  Furthermore
only a fraction of these will have properties consistent with
astronomical source activity.

In practice we find many more laboratory triggers than can be expected
from purely statistical fluctuations.  This is because the time series
we are searching are not Poissonian.  They are dominated by the
activity of real sources.  Astronomical objects, the sun, terrestrial
photon sources, and the interaction of particles in the detectors
contribute to the count rates.  Figure \ref{fig:bkgplot} shows an
integral distribution of $S_d(k)$ for a single detector ($d = 0$)
taken from a search of one day of data on the 1.024 s time scale.  For
comparison, the expectation from a normal distribution with zero mean
and unit variance is also shown.  The excess over what is expected
from a normal distribution reflects both deficiencies in the
background estimation and the activity of real sources, although no
laboratory triggers were detected by our off-line search in this
detector on the day shown.

\section{Sensitivity}
\label{sec:sensitivity}
The search strategy described above is expected to detect bursts that
were fainter than those detected by the onboard trigger mechanism.  In
this section, we estimate the efficiency of the search algorithm for
detecting events with certain physical characteristics.  In general
the ability of our search strategy to detect an event depends on its
peak flux (as measured on the search time scale), its time profile,
the background levels in the detectors, and the spacecraft orientation
with respect to the source direction.  Here we will estimate the
trigger efficiency and sky exposure of our off-line search strategy
assuming an event profile for which a single time bin completely
determines our ability to detect the event.  The effects of a more
complicated time profile are then considered separately for a
simplified case of ``slow-rising'' events that would have biased the
onboard background estimate.

\subsection{Trigger Efficiency}
\label{sec:trigeff}
We define the trigger efficiency $E(P,\nu,\alpha,\delta,t)$ to be the
probability that an event with given physical characteristics will
satisfy the laboratory trigger criteria.  The event is modeled in
terms of its peak flux on the time scale of the search ($P$),
power-law photon spectral index ($\nu$), source direction in
equatorial coordinates $(\alpha, \delta)$, and time of occurrence
during the mission ($t$).  The background rates in the detectors, the
spacecraft orientation, and the geographic position of the spacecraft
must also be known to estimate $E$, but these quantities are known
from the data once $t$ is specified.  The time profile of the event is
considered to be a square pulse that occupies a single time bin.

For a particular spacecraft orientation and position, the mean count
rates expected from an event with given values of $P$, $\nu$,
$\alpha$, and $\delta$ can be computed by adding the ``direct'' count
rates found from the BATSE instrument response matrix to the
``scattered'' count rates expected from the scattering of incident
photons by the Earth's atmosphere.  The instrument response matrices
and atmospheric scattering model are described further in Pendleton et
al. (1995) and Meegan et al. (1996).\footnote{The detector response
  and atmospheric scattering matrices used in this work were provided
  by the BATSE team.  We also made use of some elements of the
  BACODINE burst location code (Scott Barthelmy, private
  communication) which is based on an early version of the BATSE
  LOCBURST code.} The resulting total count rates are multiplied by
the time binning interval to obtain the expected counts (above
background) in a single bin of the time series for each detector; we
denote these quantities by $C^{*}_d$.  The expected background counts,
$B_d$, are estimated from the measured rates at time $t$ and the
expected significances in the detectors are calculated as $S^{*}_d =
C^{*}_d/\sqrt{B_d}$.  Let $s^{*}_1$ and $s^{*}_2$ be the the greatest
and second greatest of the $S^{*}_d$, respectively.

The trigger efficiency, $E$, is the probability that a measurement of
a pair $(s_1, s_2)$ will meet the laboratory trigger criteria.  The
counting statistics imply that our measurement, $(s_1, s_2)$, is drawn
from a bivariate normal distribution with unit variances centered on
the expected significances, $(s^{*}_1, s^{*}_2)$.  The probability
that we detect the given event is therefore estimated by integrating this
distribution over the area, $A$ (shown in Figure \ref{fig:trigarea}),
in which the trigger criteria are satisfied:
\begin{equation}
E = \int\!\!\int_{A} \frac{1}{2 \pi} \exp \left[ -\frac{(s_1 -
    s^{\ast}_1)^{2}}{2} \right] \exp \left[ -\frac{(s_2 -
    s^{\ast}_2)^{2}}{2} \right] \,ds_1 \,ds_2.
\end{equation}
We evaluate $E$ on a three-dimensional grid composed of 9 peak fluxes,
252 source directions, and 4992 times per orbital precession period of
\cgro\/\ (i.e., every 15 minutes).  The photon spectral index is fixed
at $\nu = 2.0$.  The 9 peak fluxes were chosen to span an intensity
range where the efficiency varies significantly.  The 252 source
directions are nearly isotropically distributed on the unit sphere
(\cite{Tegmark96}).  The 4992 times per orbital precession period were
chosen to thoroughly sample the range of background variations.  For
points where the source direction is behind the Earth or no searchable
data are available, $E$ is set to zero.  This calculation must be
repeated for each time scale and energy channel combination in our
search.

Figure \ref{fig:teff} shows $E$ as a function of $P$ for events
searched on the 1024 ms and 4096 ms time scales in the 50--300 keV
range.  The source directions and times of occurrence have been
averaged.  For comparison, the trigger efficiency on the 1024 ms
time scale from the BATSE 1B catalog is also shown (\cite{Batse1b}).

For the 1024 ms time scale, Figure \ref{fig:teff} shows that our
search is nearly complete near the BATSE threshold ($\sim$ 0.2 ph
cm$^{-2}$ s$^{-1}$).  Our off-line search should detect about 50\% of
events with peak fluxes lower by a factor of $\sim 2$ than the onboard
50\% completeness limit.  For events that maintain their peak flux for
at least 4 s or 8 s, the searches on the 4096 ms and 8192 ms time
scales can reach even lower peak fluxes.  The values of $E(P)$ shown
in Figure \ref{fig:teff} vary by only a few percent between spacecraft
orbital precession periods (52 days for \cgro).

The sky coverage of our search is determined by the angular
distribution of $E(P,\nu,\alpha,\delta)$, where $P$ and $\nu$ are
fixed and $t$ has been averaged.  If $T$ is the total time period
covered by the data searched, then $T \times E(P,\nu,\alpha,\delta)$
gives the total amount of time that our search was sensitive to an
event with the given intensity, spectral index, and source direction.
In practice our search covers many orbits, so the dependence on the
equatorial right ascension ($\alpha$) averages out.  Figure
\ref{fig:skyexp} shows the sky exposure as a function of declination
($\delta$) for events with $P \ge 0.5$ ph s$^{-1}$ cm$^{-2}$ on the
1024 ms time scale and $\nu = 2.0$.  The decreased exposure near the
celestial equator is due to Earth blockage.  The Southern Hemisphere
gets less exposure than the Northern Hemisphere due to passages
through the SAA.

\subsection{Sensitivity to ``slow-risers''}
\label{sec:slowsens}
The chances that our search detects a given event also depend on its
time profile in our time series.  For example, an event with a peak
flux near the detection threshold that is shaped like a square pulse
occupying $N$ time bins will have $N$ statistical chances to meet or
exceed the trigger criteria.  In such a case the trigger efficiency
calculated above will be an underestimate.

On the other hand, the event profile may be such that it biases the
nominal background rate used by the search algorithm and artificially
raises the detection threshold.  Then the trigger efficiency
calculated above is an overestimate.  This case is more serious as it
implies that the search could miss a population of such events even
though their peak flux is well above the nominal minimum detection
threshold.  Both the onboard trigger mechanism and our off-line search
are subject to this limitation.

The most problematic time profile is one which both rises and decays
slowly on a time scale long compared to the background averaging with
no significant rapid variability.  This case is difficult to
distinguish from background variations arising from the spacecraft
environment.  Neither our search nor the onboard trigger has
appreciable sensitivity to events of this type.  They are even
unlikely to be evident in a close visual inspection of the count
rates.

An event that rises slowly and then either falls off quickly or
subsequently goes into a more complicated profile can usually be
identified as a transient, however.  We will call such an event a
``slow-riser.''  The onboard trigger mechanism has particular trouble
with slow-risers because it can only base its background estimate on
count rates measured over some 17.408 s interval that occurred some
time during the 34.8 s before the time bin being tested.  Lingenfelter
\& Higdon (1996) have discussed this effect for the onboard burst
trigger.  Here we estimate the extent to which our off-line search
algorithm is sensitive to slow-risers.

The profile of an idealized slow-riser is shown in Figure
\ref{fig:slowrise}.  The event is characterized by a peak amplitude,
$s$, measured in sigmas above background, and a slope, $r$, measured
in sigmas per second, where one sigma is one standard deviation of the
expected background fluctuations.  The total duration of the event is
$s/r$.  To estimate our sensitivity to such events, we generated a
grid of peak amplitudes and a grid of slopes.  For each peak amplitude
and slope, we generated 5000 events with a background level of 1000
counts s$^{-1}$ and Poisson noise to mimic the counting statistics.  We then
used our detection algorithm to find the fraction of events that met
our off-line trigger threshold and the fraction of events that met the
onboard trigger threshold.  

The results are shown in Figure \ref{fig:slowcontour}, which shows
contours of detection probability for (idealized) events with maximum
significance, $s$, and slope, $r$, when searching on the 1024 ms time
scale.  Evidently our laboratory detection algorithm is more sensitive
to slow-risers than the onboard trigger mechanism.  This is due both
to the lower detection threshold and to the use of data before and
after the time bin being tested when forming the background estimate.
Slow-risers with no subsequent variability that are both longer than
about 30 s and fainter than 5$\sigma$ (about 0.3--0.5 photons
cm$^{-2}$ s$^{-1}$ in the 50--300 keV range) are unlikely to be found
by our 1024 ms off-line search.  The 4096 ms and 8192 ms searches are
sensitive to longer events, however.

The idealized profile used for these estimates may not be
representative of the faintest or longest transients since the typical
profiles of those events are not known a priori.  The simple case
presented here shows that our search algorithm is more
sensitive to {\it some} slow-rising bursts.

\section{Classification and Analysis of Events}
\label{sec:class}
We visually inspect each laboratory trigger flagged by our off-line
search to separate the events into useful categories.

We first examine plots of the count rates in each of the eight
detectors at the time of the laboratory trigger.  This information is
adequate to exclude from further analysis the majority of features in
the data that are not interesting in the context of our search.
Common examples of such features are occultation steps, phosphorescence
spikes, and magnetospheric particle precipitation events.

Occultation steps due to bright sources (such as Cyg X-1) rising above
(or setting below) the Earth's limb typically appear as sustained
increases (or decreases) in the count rates in two or more detectors.
This signature is not generally consistent with the burst-like
transients we seek.

Phosphorescence spikes due to the interaction of high-energy particles
in the detectors typically appear in the DISCLA data as short ($<$ 1.024
s), intense ($>$ 1000 counts s$^{-1}$) spikes in the lowest energy
channel (25--50 keV) of a single detector.  The intense signal in one
detector with no coincident signal in a second detector is
inconsistent with a point source of photons.

Magnetospheric particle precipitation events occur when particles
(mostly electrons with energies of tens of eV to tens of MeV) that are
usually trapped in the radiation belts by the Earth's magnetosphere
are released into the upper atmosphere (\cite{Burgess}). The LADs
detect the bremsstrahlung generated as precipitating particles
interact in the atmosphere or in the spacecraft (\cite{Horack} and
references therein).  These events can appear in the data in three
different ways as discussed by Horack et al. (1992).  Events of the
first kind show a smooth rise and decay with comparable intensities in
all eight detectors; this signature is inconsistent with a point
source of photons.  The second kind of event arises when the
precipitation occurs at some distance from the spacecraft, so that it
appears only in the detectors on one side of the satellite.  Because
the source of radiation is relatively nearby, however, the orbital
motion of \cgro\ gives the event different profiles in different
detectors.  The third kind of event shows rapid variability with
complex temporal structure, and can closely mimic a GRB.  Such events
can be recognized if they show characteristics from the first two
classes, such as appearance in opposite-facing detectors or
inconsistent time profiles between detectors.  These events also
exhibit a bremsstrahlung cutoff in their energy spectrum.

If an event does not appear to be any of the above, we define
background intervals by hand and estimate the source direction using
our version of the BATSE LOCBURST software (see below).  We also
examine the profile of the event in the 4 DISCLA energy channels.

Events with directions estimated to be behind the Earth are classified
as ``Earth'' events.  We have not yet done a detailed analysis of this
category, but it is likely to include electron precipitation events
occurring below the spacecraft, events from other categories for which
we obtained poor direction estimates, and possibly terrestrial gamma
ray flashes (TGRFs; \cite{Fishman94}).

Solar flares are identified based primarily on spectral softness and
location.  Typical solar flares have most of their counts in DISCLA
channel 1, with fewer in channel 2, and fewer still in channel 3.  In
contrast, typical GRBs have most of their counts in channels 2 and 3,
with proportionally less signal in channel 1.  Some spectrally hard
solar flares are observed, however, so events with directions
consistent with the sun (within uncertainties) are classified as
solar.  For the search described here we did not compare laboratory
trigger times with records of solar activity to separate hard solar
flares from GRBs.

Events which appear to be neither terrestrial nor solar and which have
sufficient spectral hardness to be seen both in channels 2 and 3 are
classified as GRB candidates.

Events which do not make it into any of the previous categories are
classified as ``unknown'', a category which includes all the
low-energy (channel 1 only) events that are not obviously of
terrestrial or solar origin.  Events with significant counts in
channels 1 and 2 (25--100 keV) but not in channel 3 (100--300 keV) are
also included in this category.

The classification of laboratory triggers is subjective in cases where
there is not an obvious indication of the nature of the event.  To
compare our classification methods with those of the BATSE team, we
can use the 317 onboard triggers that were also laboratory triggers in
our search of 345 days of data (see section \ref{sec:results}; the
remaining 256 triggers were not detected because of gaps in the DISCLA
data files).  Table \ref{tbl:ClassMat} shows how these onboard
triggers were classified by the BATSE team and by us.  

Out of the 221 events that were identified as GRBs by the BATSE team,
we classified 198 as GRBs, 10 as solar flares, 4 as magnetospheric
events, and 9 in other categories (such as ``unknown'' or ``Earth'').
The 10 events that were classified as GRBs by the BATSE team but as
solar flares by us all had estimated directions consistent with the
sun.  We classified them as solar flares because we could not argue
that they were GRBs rather than hard solar flares.  The 4 BATSE GRBs
that we classified as magnetospheric events occurred while the
spacecraft was in the vicinity of the SAA or at a maximum (or minimum)
geographic latitude, where particle precipitation events are common.
The remaining 9 BATSE GRBs that we put in other categories include
very short (duration $<$ 1 s) bursts that were difficult to classify
due to low signal-to-noise, events for which we estimated the
directions to be behind the Earth, and bursts for which there was an
unusually low count rate in channel 3 (100-300 keV).

The results in Table \ref{tbl:ClassMat} show that our classifications
agree with those of the BATSE team in most cases.  When there is
uncertainty between a GRB and a non-GRB origin, our classification is
``conservative'' in that we tend towards the non-GRB classification.
There were only 2 events out of 200 that we classified as GRBs but
that the BATSE team classified as magnetospheric particle
precipitation events.  Over the course of a year, our tendency to
classify hard events occurring near the sun as solar flares introduces
a bias against GRBs that occur in the plane of the ecliptic.

For each of the solar flares, GRB candidates, and unknown events we
estimate a source direction, intensity, and power-law spectral index
using a modified version of the BATSE LOCBURST code.  This software
uses the BATSE detector response matrices along with a model for the
scattering of incident photons by the Earth's atmosphere to find the
count rates expected from an event with intensity $P$ (in photons
cm$^{-2}$ s$^{-1}$ above 10 keV), power-law spectral index $\nu$, and
source direction $(\theta, \phi)$ in \cgro\ coordinates
(\cite{Batse1b}; \cite{Pendleton95}).  Let $C^{i}_{d}(P, \nu, \theta,
\phi)$ denote the model count rates expected in energy channel $i$ of
detector $d$, and let $\tilde{C}^{i}_{d}$ denote the measured
(background subtracted) count rates with associated statistical
measurement uncertainties $\tilde{\sigma}_{i,d}$.  To estimate
$\tilde{P}$, $\tilde{\nu}$, $\tilde{\theta}$, and $\tilde{\phi}$ for a
given event, the software minimizes the following measure of
goodness-of-fit,
\begin{equation}
\chi^2(\tilde{P},\tilde{\nu},\tilde{\theta},\tilde{\phi}) =
\sum_{i,d} \left[\frac{\tilde{C}^{i}_{d} -
    C^{i}_{d}(\tilde{P},\tilde{\nu},\tilde{\theta},\tilde{\phi})
  }{\tilde{\sigma}_{i,d}}\right]^2.
\end{equation}
The measured count rates used for this procedure are mean count rates
obtained from the main (most intense) portion of the burst as selected
by hand during our visual inspection of the laboratory triggers.

The angular response of the BATSE detectors can allow for multiple,
widely separated local minima in the $\chi^2$ parameter space,
especially for weak bursts.  For example, if a burst has most of its
counts in just two detectors it can be equally consistent with two
burst directions depending on the choice of the third most brightly
illuminated detector.  For weak bursts, the statistics may not be
good enough to reliably distinguish the third detector.  Background
variations (due to real source activity) in some detectors can also
make it difficult to distinguish the third detector.  In such 
situations, $\chi^2$ can be a strongly non-linear function of the
observed rates.  The errors on the estimated burst parameters cannot
then be reliably estimated from the formal covariance matrix of the
fit.

To estimate errors on the model parameters for each burst, we produce
50 sets of ``synthetic'' burst rates obtained by drawing from a random
distribution with the same means and variances as the measured rates.
These 50 sets of synthetic count rates are then subjected to the same
$\chi^2$ minimization procedure as the real rates.  The variances in
the parameters obtained from the synthetic count rates are used to
estimate the uncertainty on the parameters ($\tilde{P}$,
$\tilde{\nu}$, $\tilde{\theta}$, $\tilde{\phi}$) obtained from the
measured count rates.

Using the estimated mean intensity $\tilde{P}$ we obtain a conversion
factor from counts s$^{-1}$ in the most brightly illuminated detector
to units of ph cm$^{-2}$ s$^{-1}$ in the 50--300 keV range.  The peak
flux and fluence of each event in physical units are then determined
by multiplying the corresponding measured counts by this conversion
factor.  The durations of events are characterized by the $T_{50}$ and
$T_{90}$ duration measures, which are the time intervals during which
the burst fluence increases from 25\% to 75\% and from 5\% to 95\%
(respectively) of the total fluence (\cite{Kouveliotou93};
\cite{Koshut96}).  Uncertainties in the peak flux, fluence, and
durations are derived from the uncertainties in $\tilde{P}$,
$\tilde{\nu}$, and the measured (background subtracted) count rates
using the standard techniques for the propagation of small random
errors (although the assumptions required by this method are not
always satisfied).

We have attempted to ascertain how well the above procedures estimate
burst intensities and directions by applying our methods to GRBs from
the 3B catalog (\cite{Batse3b}).  Because the burst intensities and
directions must be estimated simultaneously by folding a model through
the detector and atmospheric response matrices, systematic errors in
the inferred quantities can arise from the background subtraction, the
modeling of the event spectrum, spectral changes during the event, and
detector calibration.

To evaluate the accuracy of our intensity measurements, we applied our
analysis procedure to 29 GRBs from the 3B catalog (\cite{Batse3b}).
We compared our peak fluxes (derived from the DISCLA data) to those
obtained by the BATSE team (using the high resolution burst data
types; see \cite{Pendleton96}).  In 13 bursts the two measurements
agreed to within the 1$\sigma$ statistical uncertainties, and in 24 bursts
they agreed to within 2$\sigma$.  The largest
disagreements (in terms of standard deviations) occur only in the most
intense bursts, where the systematic errors are expected to dominate
the statistical uncertainties.  In those cases, the measurements
disagree by less than 30\%. 

At least two sources of systematic errors are responsible for the
differences between our peak flux measurements (on intense bursts) and
those obtained by Pendleton et al. (1996) for the 3B catalog.  First,
we model the incident photon spectrum as a power law, whereas
Pendleton et al.  (1996) allow for some curvature in the spectrum.
Second, our peak flux measurement on the 1024 ms time scale is based
on the DISCLA time bin that contains the most counts above background.
The phase of this 1024 ms bin relative to the ``true'' 1024 ms peak
flux depends on the DISCLA sampling times; therefore, the ``true''
1024 ms peak flux could be spread over two DISCLA time bins.  In
contrast, Pendleton et al.  (1996) use data with 64 ms time resolution
to find which placement of a 1024 ms interval yields the highest peak
flux on that time scale.

A similar procedure was used to assess the accuracy of our direction
estimates.  Comparison of a sample of GRBs from the 3B catalog and
hard solar flares which triggered onboard suggests that an additional
systematic uncertainty of about 4$^{\circ}$ should be added in
quadrature to our statistical direction uncertainties for events with
emission between 50 and 300 keV.  This is expected since our version
of the LOCBURST code roughly corresponds (with minor improvements) to
the version used to produce the BATSE 1B catalog (\cite{Batse1b}).
The improvements to LOCBURST that reduced systematic errors to
1.6$^{\circ}$ for the 3B catalog require more spectral information
than is available in the DISCLA data and have not been incorporated
into the analysis described in this paper.  For the channel 1 only
(25--50 keV) events, comparison with solar flares indicates that the
additional systematic uncertainty is about 6$^{\circ}$.

For faint events the background subtraction can be a substantial
source of systematic error because it is not always clear what is
background and what is low-level emission before or after the event.
The duration estimates (and thus fluence estimates) are particularly
sensitive to the choice of background intervals (see \cite{Koshut96}).

\section{Results}
\label{sec:results}
We have applied the search and analysis procedures described in
sections \ref{sec:searchalg} and \ref{sec:class} to the DISCLA data
taken between 1993 January 13 and 1993 December 24.  The corresponding
range of Truncated Julian Day numbers is TJD 9001--9345 (TJD = Julian
Day $-$ 2,440,000.5).  Figure \ref{fig:allevents} shows a sky map
which combines events from the GRB candidate ($\diamond$), solar flare
($\ast$), and unknown (+) categories.  The concentration of solar
events in the ecliptic plane is clearly visible, as is a concentration
of low-energy (+) events in the vicinity of Cyg X-1
($\ell=71.3^{\circ}$, \mbox{$b=3.0^{\circ}$}).

\subsection{Gamma ray burst candidates}
Our search so far yields 91 non-triggered GRB candidates.  Tables
\ref{tbl:grbs1} and \ref{tbl:grbs2} together are a catalog of these
events.  

The first column of Table \ref{tbl:grbs1} is a name which specifies
the approximate time of the event in the format NTB {\em yymmdd.ff},
where $yy$ is the year, $mm$ is the month, and {\em dd.ff} is the day.
The second column gives the time of the laboratory trigger expressed
as the TJD and the seconds of day (SOD).  The next three columns give
the estimated source direction in equatorial (J2000) coordinates and
its associated statistical uncertainty.  The full $1.0\sigma$
uncertainty in the direction estimates is obtained by combining the
statistical uncertainty and the $4^{\circ}$ systematic uncertainty
(see section \ref{sec:class}) in quadrature.  The next column gives
the largest value (among the three time scales in Table
\ref{tbl:TriggerParameters}) of $C_{max}/C_{min}$, which is the ratio
of the maximum count rate achieved during the event to the minimum
count rate required for detection in the 50--300 keV band.  Events
with $C_{max}/C_{min} < 1.0$ were detected in the ``catch-all'' search
of the 25--300 keV band (search ``c'' of Table
\ref{tbl:TriggerParameters}) and may represent spectrally softer GRBs
that had too few counts to trigger in the 50--300 keV band.  The next
column gives the threshold number of counts in the 50--300 keV band,
$C_{min}$, and (in superscript) the search time scale which yielded
the largest value of $C_{max}/C_{min}$.  The next column lists the
searches from Table \ref{tbl:TriggerParameters} in which the event was
detected.

The last column of Table \ref{tbl:grbs1} gives the reasons in our
estimation why the events did not trigger onboard the spacecraft.  The
notation ``R'' indicates the event occurred during the readout of a
previous onboard trigger. ``D'' indicates that the event occurred
while the onboard trigger was disabled due to passage through a region
with a high probability of magnetospheric particle precipitation
events.  ``F'' indicates that the event was too faint to meet the
onboard burst trigger threshold. ``BB'' indicates that the event
failed to trigger onboard because it biased the onboard background
average.  In one case, no experiment housekeeping data were available
so the state of the onboard trigger at the time of the event could not
be determined.  A ``?'' is entered for this event because we cannot
determine whether the reason is ``D'' or ``BB''.  (See below for
further discussion of these reasons and examples of events in each
category.)

Table \ref{tbl:grbs2} gives the durations and intensities (in physical
units) of the GRB candidates.  The first column gives the name of the GRB
candidate.  The next two columns give estimates of the $T_{50}$ and
$T_{90}$ duration measures.  Events with no entry in the $T_{50}$
column had their $T_{90}$ duration estimated by eye.  Uncertainties
listed as $0.00$ indicate that the uncertainty is less than the
duration of one DISCLA time bin (1.024 s).  The next column gives the
peak flux in the 50--300 keV range as measured on the 1024 ms time
scale.  The next column gives the 50--300 keV fluence estimate.

Figure \ref{fig:grbsky} shows a sky map of the direction estimates for
the 91 untriggered GRB candidates.  Events are shown as 1.0$\sigma$
error circles centered on the best-fit location.  Using the sky
exposure calculated in section \ref{sec:trigeff}, the dipole and
quadrupole moments of this direction distribution in Galactic
coordinates are $\langle \cos\theta \rangle = -0.001 \pm 0.025$ and
$\langle \sin^{2}b -1/3 \rangle = -0.022 \pm 0.021$, where $\theta$ is
the angle between the burst direction and the Galactic center and $b$
is the Galactic longitude.  These values are consistent with the
values $\langle \cos \theta \rangle = 0.00 \pm 0.06$ and $\langle
\sin^{2}b - 1/3 \rangle = 0.00 \pm 0.03$ expected from an isotropic
distribution with the same number of bursts.

The dipole and quadrupole moments (corrected for sky exposure) with
respect to equatorial coordinates are $\langle \sin\delta \rangle =
0.036 \pm 0.027$ and $\langle \sin^{2} \delta - 1/3 \rangle = 0.074
\pm 0.024$.  The dipole moment is consistent with that expected from
an isotropic distribution, $\langle \sin \delta \rangle = 0.00 \pm
0.06$.  The quadrupole moment appears to be only marginally consistent
with the value $\langle \sin^{2} \delta - 1/3 \rangle = 0.00 \pm 0.03$
expected for an isotropic distribution, indicating a weak
concentration of events in the direction of the celestial poles.  This
result may be due to our tendency to classify GRBs with
directions consistent with the sun as hard solar flares (see section
\ref{sec:class}).

The durations based on the $T_{90}$ interval of these events range
from $\approxlt 1.024$ s to $\sim 350$ s.  We have examined the
non-triggered GRB candidates to see if any appear to be related to an
onboard triggered GRB that occurred within 1 day of a non-triggered
event.  In a combined sample of 91 non-triggered GRB candidates and
333 bursts from the 3B catalog (covering TJD 8995--9347), we found 7
pairs of bursts occurring within 1 day of each other and having
direction measurements compatible within 1$\sigma$ uncertainties.
Only 3 of these 7 pairs involved a non-triggered GRB candidate.  We do
not consider this to be evidence that any of these pairs share the
same burst source.  We expect statistically to find 5--8 such pairs in
a sample of the same size drawn from bursts randomly distributed
uniformly in time and isotropically in space, with median location
measurement errors of 8--10$^{\circ}$.  If one or more convincing
pairs had been identified, however, they could have either been
interpreted as burst repetition (see \cite{Wang94}) or as bursts that
extend knowledge of the $T_{90}$ distribution to longer durations.

Figure \ref{fig:pflux} shows the integral peak flux distribution on
the 1024 ms time scale for the untriggered GRB candidates.  These
events are concentrated at the faint end of the distribution, as
expected for events which were generally too faint to cause an
onboard trigger.  When the non-triggered bursts are added to the
triggered bursts for the same time period, the departure from the
$-3/2$ power law slope expected from a homogeneous distribution (in
Euclidean space) remains evident.  

The top two panels of Figure \ref{fig:slnlp} show the differential,
$n(P)$, and integral, $N(>P)$, distributions of the peak fluxes, $P$,
for the combined sample of 83 non-triggered GRB candidates and 233
onboard triggered GRBs that were detected on the 1024 ms time scale.
The bottom panel shows the slope of the logarithmic number versus peak
flux distribution, defined by
\begin{equation}
s(P) = \frac{d \log N(>P)}{d \log P} = - P \frac{n(P)}{N(>P)}.
\end{equation}
The dotted histograms in Figure \ref{fig:slnlp} show the distributions
corrected for our laboratory trigger efficiency (see figure
\ref{fig:teff}).  The logarithmic slope $s(P)$ is consistent with a
value of $-0.5$ $\pm$ 0.1 at peak fluxes of 0.15--0.35 ph cm$^{-2}$
s$^{-1}$.

Another measure of the inhomogeneity of the source distribution is
reflected in the distribution of the $V/V_{max}$ statistic for these
events, given by $(C_{max}/C_{min})^{-3/2}$ (\cite{Schmidt88}).  For
the 91 non-triggered GRB candidates, the average value $\langle
V/V_{max}\rangle = 0.28 \pm 0.03$. This value is biased by the
elimination of the strong (triggered) bursts; but it is of interest
when considering how biases inherent in the onboard trigger mechanism
could affect conclusions about the spatial inhomogeneity of GRB
sources.  For example, if the onboard trigger's bias against
slow-risers had significantly biased the value of $\langle V/V_{max}
\rangle$ = 0.33 $\pm$ 0.01 obtained for the 3B catalog
(\cite{Batse3b}) then the value obtained from the non-triggered GRB
candidates alone could be expected to be much higher.  For comparison,
the value obtained using the cataloged bursts detected during the same
time period is $\langle V/V_{max}\rangle = 0.13 \pm 0.02$.  For the
{\it combined} non-triggered GRB candidates and cataloged bursts,
$\langle V/V_{max}\rangle = 0.18 \pm 0.02$.\footnote{These values are
  corrected from the erroneous values that appeared in earlier
  versions of this paper and in the version that appeared in ApJ, 491,
  704 (1997).  See erratum at end of paper.}

Figure \ref{fig:srplot} shows intensity profiles of some
representative GRB candidates.  Two adjacent plots are shown for each
event.  The plots on the left show the burst profile from the detector
most brightly illuminated by the burst.  The plots on the right show
the profile from the second most illuminated (or ``second brightest'')
detector; they illustrate why some of the events did not cause an
onboard burst trigger.  The dashed lines show our estimate, $B^{\rm
  fit}_k$ of the background counts in each bin ($k$) based on
polynomial fits to data before and after the event.  The dotted lines
show the 5.5$\sigma_B$ threshold level given by $5.5\sqrt{B^{\rm
    fit}_k}$.  This level represents an ``ideal'' threshold estimate
and it is in general different from the actual onboard background
estimate that was in effect at the time the event occurred (see
section \ref{sec:searchalg}).  The onboard background estimate has a
statistical uncertainty resulting from the uncertainty in the mean
count rate during the 17.408 s background accumulation interval.  The
dot--dashed line represents the threshold corresponding to
5.5$\sqrt{B^{\rm fit}_k}$ {\it plus} the uncertainty in the onboard
trigger level arising from the onboard background uncertainty.  An
event with peak counts just slightly above our ``ideal'' 5.5$\sigma_B$
threshold may fall below the onboard burst trigger threshold even
though both thresholds are based on background estimates that are
statistically consistent with each other.  We classify events with
peak counts less than 5.5$\sqrt{B^{\rm fit}_k}$ plus the onboard
threshold uncertainty as too faint to trigger onboard (``F'' in table
\ref{tbl:grbs1}).

The event in row (a) of Figure \ref{fig:srplot} failed to trigger
onboard because it occurred during the read out period of a previous
onboard burst trigger.  The event in row (b) occurred while the onboard
burst trigger was disabled due to spacecraft passage through a region
identified with a high probability of a false trigger due to
atmospheric electron precipitation events.  The events in rows (c) and
(d) failed to trigger onboard because the onboard background estimate
was biased upwards by slowly rising burst flux; these are examples of
slow-risers.  Panels (e) and (f) show events that were too faint to
meet the onboard trigger threshold.

We estimate that the 91 GRB candidates failed to trigger onboard the
spacecraft for the following reasons: 15 events occurred during the
read out of a brighter event, 2 occurred while the onboard trigger was
disabled for other reasons, 63 were below the $5.5 \sigma_B$ threshold
in the second brightest detector, and 10 had a slow rise that modified the
onboard background estimate.  One occurred during a time for which no
spacecraft housekeeping data are available to determine the status of
the onboard trigger.

The onboard trigger mechanism's bias against slow-rising GRBs has been
discussed by Lingenfelter and Higdon (1996).  The 10 (possibly 11)
events we find that failed to trigger onboard the spacecraft {\it
  solely} because of the slow-rising effect constitute 3.0\%
(possibly 3.3\%) of the total 332 GRBs that have been detected above
the onboard threshold while the trigger was active.  This is a lower
fraction than estimated elsewhere (\cite{Ling96}).  We note, however,
that our search algorithm is biased against faint events which rise on
time scales longer than $\sim 30$ s on the 1.024 s time scale (see
Figure \ref{fig:slowcontour}).

\subsection{Unknown events}
The ``unknown'' category of laboratory triggers includes all events
which were not obviously of terrestrial or solar origin and which do
not resemble a GRB.  Most of the channel 1 only (low-energy, 25--50
keV) events fall into this category.  The major problem with this
class of events is that it is dominated by intensity fluctuations from
Cyg X-1: of 799 events in the unknown category, 689 are consistent
with this source (although they may not {\it all} be from Cyg X-1);
see the clustering of events marked (+) in Figure \ref{fig:allevents}.
If we remove all the events consistent with Cyg X-1 we are left with
the sky map shown in Figure \ref{fig:unksky}, where events are plotted
as their $1.0 \sigma$ error circles.  Although Figure \ref{fig:unksky}
shows some general clustering toward the galactic center, there is no
obvious clustering that would indicate the activity of any particular
source.

We find two events which can convincingly be attributed to SGR
1806$-$20 based on intensity, spectral softness, and location.  Both
occur within one day of the onboard triggered emission from SGR
1806$-$20 reported by Kouveliotou et al. (1994).  Recent activity from
this source suggests that more events from SGR 1806$-$20 (or other
SGRs) may be detectable when this search is extended
(\cite{Kouveliotou96}).

\section{Conclusions}
Our search of 345 days of archival BATSE data has uncovered a
significant number of astronomically interesting transients.  

The 91 non-triggered GRB candidates detected (so far) by this search
include some of the faintest GRBs ever observed.  When combined with
the bursts detected by the onboard trigger during the same 345 days,
these events extend knowledge of the peak flux distribution to values
a factor of $\sim$ 2 lower than the onboard detection threshold.  Near
the onboard trigger threshold, the combined sample is expected to be
nearly complete (on the 1024 ms time scale). We find the logarithmic
slope of the integral number versus peak flux distribution to be $-0.5
\pm 0.1$ at peak fluxes of 0.15--0.35 ph cm $^{-2}$ s$^{-1}$ after
correcting for our laboratory trigger efficiency.  The value of
$\langle V/V_{max} \rangle = 0.18 \pm 0.02$ for the combined
sample.\footnote{This value is corrected from the erroneous value that
  appeared previously.  See erratum at end of paper.}  We find no
evidence for anisotropy in the direction distribution of these events.

These results are consistent with those obtained from analyses of the
published BATSE catalogs.  The biases inherent in the onboard
trigger mechanism do not appear to have significantly undermined its
sampling of GRBs, at least for bursts with the characteristics our
search can detect.

The non-triggered GRB candidates add to the database of GRBs available
for duration studies, searches for burst repetition, and searches for
gravitational lensing (see \cite{FishmanReview95} for an overview and
references).  The slow-risers and the more intense bursts (which
occurred while the onboard trigger was disabled) will probably be the
most useful non-triggered GRBs for such purposes.

The low-energy (25--50 keV) events detected by our search arise from a
variety of sources.  While intensity fluctuations from Cyg X-1
dominate this class of events, we find a significant number that must
be due to other sources.  Because of the difficulty in accurately
estimating the source directions of these events, identification of
the individual sources responsible for them depends on unique
repetition patterns or temporal coincidences with other observations
(as in the case of events from SGR 1806$-$20).  The possibility
remains that we may identify among these events new source activity or
completely new burst sources.

The effort to extend this search to cover the more than 5 years of
remaining archival data is in progress.

\acknowledgments J.\ M.\ K.\ acknowledges support from a National
Science Foundation Graduate Research Fellowship during the preliminary
phase of this research and subsequent support from NASA Graduate
Student Researchers Program Fellowship NGT8-52816.  W.\ H.\ G.\ L.\ 
acknowledges support from NASA under grant NAG5-3804. C.\ K.\ 
acknowledges support from NASA under grant NAG5-2560. J.\ v.\ P.\ 
acknowledges support from NASA under grant NAG5-2755.  We also thank
Scott Barthelmy and James Kuyper for sharing the BACODINE version of
the BATSE LOCBURST code, which formed the foundation for our trigger
efficiency and burst location algorithms.

\clearpage

\begin{center}
ERRATUM
\end{center}

In the paper ``A Search for Nontriggered Gamma-ray Bursts in the BATSE
Data Base'' by Kommers et al. (ApJ, 491, 704, [1997]) the values of
$\langle V/V_{max}\rangle$ that include onboard-triggered bursts are
incorrect.  Owing to a programming error, whenever a value of
$C_{min}$ was available for a burst listed in the 4B catalog, that
catalog value (appropriate for the onboard burst trigger) overwrote
the value appropriate for our more sensitive off-line search.  Thus
the values of $V/V_{max}$ were overestimated by a factor of $\sim
2^{3/2}$ for the bursts that were triggered onboard.  This error does
{\it not\/} affect the values of $C_{max}/C_{min}$ and $C_{min}$ for
the non-triggered GRBs listed in Table 3, nor does it affect the value
of $\langle V/V_{max}\rangle = 0.28 \pm 0.03$ computed for the
non-triggered (only) GRB sample.

The correct value of $\langle V/V_{max}\rangle$ for the
onboard-triggered bursts detected by our off-line search is $\langle
V/V_{max}\rangle = 0.13 \pm 0.02$.  For the {\it combined}
non-triggered GRB candidates and onboard-triggered bursts, $\langle
V/V_{max}\rangle = 0.18 \pm 0.02$.  This value is significantly lower
than the value of $\langle V/V_{max}\rangle = 0.33 \pm 0.01$ obtained
for the 3B catalog.  These corrections, therefore, significantly {\it
  strengthen\/} the conclusion of the paper, that the BATSE onboard
trigger has {\it not\/} missed a significant population of faint
gamma-ray bursts owing to the biases inherent in the onboard burst
trigger mechanism.

\clearpage


\clearpage


\begin{figure}
\caption{Schematic diagram of the laboratory (off-line search) trigger criteria.
  If the significances of the fluctuations measured in the two most
  brightly illuminated detectors lie in the non-shaded area $A$, our
  search will flag that time bin as a laboratory trigger.}
\label{fig:trigarea}
\end{figure}

\begin{figure}
\caption{Integral distribution of $S_d(k)$ for the data of May 27,
  1993 (TJD 9135) in detector 0.  The dotted line shows the
  expectation for a Gaussian distribution with zero mean and unit
  variance.  The excess counts are attributed both to deficiencies in
  the background estimate and to the activity of real sources.}
\label{fig:bkgplot}
\end{figure}

\begin{figure}
\caption{Off-line trigger efficiency.  The solid line shows the efficiency of our off-line search
  algorithm for detecting an event with a given peak flux in the
  50--300 keV range on the 1024 ms time scale.  The long-dashed line
  shows the trigger efficiency for our off-line search on the 4096 ms
  time scale.  The short-dashed line shows the trigger efficiency from
  the 1B catalog.}
\label{fig:teff}
\end{figure}

\begin{figure}
\caption{Sky exposure for our 50-300 keV off-line trigger search as a
  function of declination, assuming a search spanning 345 days.}
\label{fig:skyexp}
\end{figure}

\begin{figure}
\caption{Simplified profile of a ``slow-riser''.  It reaches maximum
  significance $s$ by rising with a slope $r$ given in sigmas per second.}
\label{fig:slowrise}
\end{figure}

\begin{figure}
\caption{Contours showing the probability of detecting a
  ``slow-riser'' with a given peak significance (relative to the
  ``true'' background) and a given slope.  Thick contours apply to our
  off-line search algorithm, and thin contours apply to the onboard
  burst trigger.}
\label{fig:slowcontour}
\end{figure}

\begin{figure}
\caption{Sky map of all non-terrestrial non-triggered events in
  Galactic coordinates. The GRB candidate ($\diamond$), solar flare
  ($\ast$), and unknown (+) categories are shown.  The concentration
  of solar events in the ecliptic plane is clearly visible, as is a
  concentration of low-energy (+) events in the vicinity of Cyg X-1
  ($\ell=71^{\circ}.3$, \mbox{$b=3^{\circ}.0$}).}
\label{fig:allevents}
\end{figure}

\begin{figure}
\caption{Sky map of 91 non-triggered GRB candidates, shown as
  1$\sigma$ error circles in Galactic coordinates.}
\label{fig:grbsky}
\end{figure}

\begin{figure}
\caption{Integral number versus peak flux distribution of 91
  non-triggered GRB candidates (dotted line).  No
  corrections for trigger efficiency have been applied.  The
  distribution for GRBs from the 3B catalog detected during the same
  time period is also shown (dashed line), as is that for the combined
  sample (solid line).}
\label{fig:pflux}
\end{figure}

\begin{figure}
\caption{Peak flux distributions for 83 non-triggered GRB candidates
  combined with the 233 onboard triggered events (1024 ms time scale)
  from the same time period.  The solid histogram shows the observed
  numbers and the dotted histogram shows the numbers corrected for
  laboratory trigger efficiency.  $n(P)$ is the differential
  distribution, $N(P)$ is the integral distribution, and $s(P)$ is the
  slope of the logarithmic number versus peak flux distribution.}
\label{fig:slnlp}
\end{figure}

\begin{figure}
\caption{Intensity profiles of selected GRB candidates.  Two panels
  are used for each event.  Those on the left represent the count
  rates observed in the detector most brightly illuminated by the
  burst.  Those on the right represent the count rates in the second
  most illuminated detector.  Dashed lines represent our background
  estimate derived from polynomial fits to data before and after the
  event.  Dotted lines represent an ``ideal'' onboard trigger
  threshold based on our background estimates; the actual onboard
  trigger threshold is in general different and is based on an average
  count rate that is recomputed every 17.408 s.  The dot-dashed line
  represents the ``ideal'' onboard threshold {\it plus} the uncertainty 
  arising from the statistical uncertainty in onboard background
  average.  The burst in row (a) occurred during the read out period
  of a more intense event, and that in row (b) occurred while the
  onboard trigger was disabled.  The bursts
  in rows (c) and (d) modified the onboard background
  average.  Rows (e) and (f) show events that were too faint cause
  an onboard burst trigger.}
\label{fig:srplot}
\end{figure}

\begin{figure}
\caption{Low-energy (25--50 keV) events plotted as 1$\sigma$ error
  circles in Galactic coordinates (those consistent with Cyg X-1
  excluded).}
\label{fig:unksky}
\end{figure}

\clearpage

\begin{deluxetable}{ccccccc}
\tablehead{\colhead{Search} & \colhead{\begin{tabular}{c}Time Bin\\Duration
 (s)\end{tabular}} &
 \colhead{\begin{tabular}{c}Energy\\Channels\end{tabular}} &
 \colhead{$M$} & \colhead{$\Sigma$} & \colhead{$N_b$} &
 \colhead{\begin{tabular}{c}Statistical Detections\\per day\end{tabular}}}
\tablecaption{Parameters of the time series formed from the DISCLA
 data (see text).
\label{tbl:TriggerParameters}}
\startdata
a & 1.024 & 1 & 2.5 & 4.0 & 20 & 0.021 \nl
b & 1.024 & 2+3 & 2.5 & 4.0 & 20 & 0.021 \nl
c & 1.024 & 1+2+3 & 2.5 & 4.0 & 20 & 0.021 \nl
d & 4.096 & 1 & 2.5 & 4.0 & 15 & 0.005 \nl
e & 4.096 & 2+3 & 2.5 & 4.0 & 15 & 0.005 \nl
f & 8.192 & 1 & 2.5 & 4.0 & 5 & 0.002 \nl
g & 8.192 & 2+3 & 2.5 & 4.0 & 5 & 0.002 \nl
\enddata
\end{deluxetable}

\clearpage

\begin{deluxetable}{ccccc}
\tablehead{~ & \colhead{\begin{tabular}{c}Off-line\\GRB
 Candidate\end{tabular}} & \colhead{\begin{tabular}{c}Off-line\\Solar
 Flare\end{tabular}} &
 \colhead{\begin{tabular}{c}Off-line\\Magnetospheric\end{tabular}} &
 \colhead{\begin{tabular}{c}Off-line\\Other\end{tabular}}}
\tablecaption{Classification matrix of 317 onboard triggers that were also
 detected by our off-line search.  Rows indicate the classification
 assigned by the BATSE team, and columns indicate the classification
 assigned by us (see text).
\label{tbl:ClassMat}}
\startdata
BATSE GRB & 198 & 10 & 4 & 9 \nl
BATSE Solar Flare & 0 & 50 & 3 & 0 \nl
BATSE Magnetospheric & 2 & 2 & 30 & 4 \nl
BATSE Unclassified & 0 & 0 & 0 & 5 \nl
\enddata
\end{deluxetable}

\clearpage

\begin{deluxetable}{rrrrrrrcc}
\scriptsize
\tablehead{\colhead{Name} & \colhead{\begin{tabular}{cc}Time\\(TJD:s)\end{tabular}} & \colhead{\begin{tabular}{cc}RA\\($^{\circ}$)\end{tabular}} &
  \colhead{\begin{tabular}{cc}Dec\\($^{\circ}$)\end{tabular}} & \colhead{\begin{tabular}{cc}Err.\\($^{\circ}$)\end{tabular}}
  & \colhead{$\frac{C_{max}}{C_{min}}$} & \colhead{$C_{min}$} &
  \colhead{\begin{tabular}{cc}Searches\\Triggered\end{tabular}}  & \colhead{\begin{tabular}{cc}Reason\\Non-triggered\end{tabular}}}
\tablecaption{Times, source directions, $C_{max}/C_{min}$, searches
  triggered, and reasons for eluding the onboard trigger for 91
  non-triggered GRB candidates.
\label{tbl:grbs1}}
\startdata
NTB 930118.74 & 9005:64425.6 & 219.3 & -32.9 &   0.5 &  31.0 & 349$^{4}$ & abcdefg   & R \nl
NTB 930211.88 & 9029:76428.9 & 285.5 &  20.0 &   7.7 &   3.5 & 425$^{8}$ & bcdefg    & F \nl
NTB 930216.63 & 9034:54956.7 &  42.5 &  -8.7 &   3.6 &   2.6 & 447$^{8}$ & bcdeg     & F \nl
NTB 930217.80 & 9035:69525.6 & 311.7 & -12.0 &  28.2 &   0.9 & 286$^{4}$ & c         & F \nl
NTB 930225.86 & 9043:75141.2 & 336.6 &  35.2 &  38.4 &   1.1 & 155$^{1}$ & abc       & F \nl
NTB 930227.83 & 9045:71910.5 & 241.2 &  81.0 &  11.0 &   2.1 & 377$^{8}$ & ceg       & F \nl
NTB 930228.85 & 9046:73728.6 &  18.6 &   2.5 &   8.3 &   2.2 & 477$^{8}$ & g         & F \nl
NTB 930302.20 & 9048:17613.4 & 214.1 &  42.5 &   6.2 &   2.4 & 483$^{8}$ & g         & F \nl
NTB 930303.65 & 9049:56726.7 &   0.0 & -50.8 &   3.3 &   3.2 & 501$^{8}$ & bceg      & F \nl
NTB 930305.70 & 9051:60923.0 & 275.2 &  59.8 &   3.5 &   3.2 & 330$^{8}$ & bceg      & F \nl
NTB 930307.54 & 9053:46677.1 &  94.2 & -15.5 &  12.7 &   1.7 & 411$^{8}$ & bc        & F \nl
NTB 930308.30 & 9054:26710.7 &  88.7 & -31.6 &   3.5 &   3.8 & 399$^{8}$ & bceg      & F \nl
NTB 930310.08 & 9056: 7334.5 & 333.5 & -57.2 &   5.0 &   2.6 & 321$^{8}$ & abcdefg   & F \nl
NTB 930315.46 & 9061:40070.8 & 232.5 & -32.1 &   3.9 &   2.0 & 395$^{8}$ & g         & F \nl
NTB 930316.74 & 9062:64295.0 & 314.1 & -87.1 &  19.8 &   1.6 & 132$^{1}$ & bc        & F \nl
NTB 930318.18 & 9064:15764.1 &  16.2 &  44.0 &   8.9 &   2.4 & 425$^{8}$ & bceg      & F \nl
NTB 930320.94 & 9066:81767.0 & 228.3 &  72.7 &  15.5 &   7.2 & 436$^{8}$ & bcefg     & R \nl
NTB 930325.65 & 9071:56254.1 &  46.9 &  46.8 &   5.8 &   2.8 & 383$^{8}$ & bceg      & F \nl
NTB 930327.46 & 9073:40594.6 & 178.7 &  -5.5 &  13.2 &   3.4 & 553$^{8}$ & bc        & F \nl
NTB 930330.91 & 9076:79132.8 & 109.2 & -54.4 &   2.0 &   1.7 & 481$^{8}$ & cfg       & F \nl
NTB 930403.84 & 9080:73239.2 & 245.8 & -59.6 &   6.2 &   3.0 & 351$^{8}$ & abcdefg   & BB \nl
NTB 930409.13 & 9086:11442.8 & 315.6 &  68.8 &   4.0 &   3.3 & 334$^{8}$ & abcdef    & BB \nl
NTB 930409.91 & 9086:78639.7 & 275.4 & -17.7 &   0.9 &  12.8 & 501$^{8}$ & abcdefg   & R \nl
NTB 930410.76 & 9087:65711.8 & 182.3 & -45.3 &   7.3 &   3.0 & 380$^{8}$ & abcefg    & F \nl
NTB 930416.56 & 9093:48602.2 & 115.6 & -21.0 &   6.2 &   3.1 & 389$^{8}$ & abcdefg   & F \nl
NTB 930417.78 & 9094:68020.4 & 190.5 &   8.3 &   6.7 &   1.7 & 351$^{8}$ & abcdf     & F \nl
NTB 930421.11 & 9098:10164.9 &  23.7 &  19.3 &   7.0 &   2.0 & 520$^{8}$ & g         & F \nl
NTB 930422.58 & 9099:50820.2 &  50.4 &  -7.5 &   2.1 &   4.9 & 315$^{8}$ & bc        & ? \nl
NTB 930424.45 & 9101:38903.4 & 230.3 & -55.7 &   0.4 &  21.2 & 523$^{8}$ & abcde     & D \nl
NTB 930424.97 & 9101:84156.0 & 254.1 &  68.6 &   3.2 &   3.5 & 332$^{8}$ & abcefg    & BB \nl
NTB 930426.48 & 9103:41832.1 &  33.2 & -81.9 &   5.4 &   4.4 & 133$^{1}$ & abcdefg   & R \nl
NTB 930427.59 & 9104:51155.6 &  60.1 &  34.3 &  11.5 &   2.2 & 453$^{8}$ & abd       & F \nl
NTB 930429.75 & 9106:65094.3 &  35.7 & -25.8 &   5.4 &   5.7 & 344$^{8}$ & bcdefg    & BB \nl
NTB 930501.34 & 9108:29834.4 & 181.4 & -32.9 &  31.7 &   1.1 & 114$^{1}$ & ab        & F \nl
NTB 930506.63 & 9113:55244.4 & 259.8 &  35.4 &   8.4 &   2.4 & 548$^{8}$ & cg        & F \nl
NTB 930508.95 & 9115:82814.6 &  82.9 &  41.4 &   7.6 &   3.4 & 494$^{8}$ & bce       & D \nl
NTB 930513.98 & 9120:85533.8 & 169.0 &  12.0 &  39.5 &   1.5 & 384$^{8}$ & bc        & F \nl
NTB 930519.39 & 9126:34288.8 & 272.4 & -24.2 &  83.8 &   1.4 & 134$^{1}$ & bc        & F \nl
NTB 930612.63 & 9150:55165.1 & 254.5 &  34.3 &   2.9 &  12.1 & 277$^{4}$ & abcde     & R \nl
NTB 930616.27 & 9154:23806.6 & 179.5 &   1.8 &   4.4 &  11.7 & 165$^{1}$ & abcde     & R \nl
NTB 930617.23 & 9155:20027.0 & 244.9 & -12.7 &   6.1 &   2.5 & 299$^{8}$ & bceg      & F \nl
NTB 930626.94 & 9164:81935.0 & 342.6 & -34.9 &   3.3 &   3.6 & 512$^{8}$ & bcdefg    & BB \nl
NTB 930630.71 & 9168:61420.2 &  70.0 &  38.1 &   6.9 &   3.0 & 438$^{8}$ & bcdeg     & F \nl
NTB 930701.62 & 9169:54302.9 & 226.4 &  39.4 &   2.5 &   4.2 & 338$^{8}$ & abcde     & F \nl
NTB 930705.64 & 9173:55983.3 & 195.3 & -59.5 &   7.2 &   2.0 & 334$^{8}$ & bcefg     & F \nl
NTB 930717.20 & 9185:18101.4 & 184.3 &  57.8 &  10.3 &   1.5 & 364$^{4}$ & bcd       & F \nl
NTB 930717.98 & 9185:85357.7 & 200.5 & -66.4 &   3.5 &   3.3 & 374$^{8}$ & abc       & F \nl
NTB 930722.84 & 9190:73297.6 & 310.5 & -48.2 &   3.5 &   2.1 & 412$^{8}$ & eg        & F \nl
NTB 930728.54 & 9196:47072.9 &  90.8 &  19.0 &  21.0 &   3.3 & 510$^{8}$ & d         & R \nl
NTB 930804.71 & 9203:61858.5 &  32.7 &  66.6 &   5.0 &   3.7 & 317$^{8}$ & bcefg     & F \nl
NTB 930811.62 & 9210:53728.4 & 347.5 &  65.0 &   3.6 &   2.9 & 325$^{8}$ & bc        & F \nl
NTB 930812.27 & 9211:23904.9 & 198.0 & -27.8 &   2.4 &   4.9 & 409$^{8}$ & bcdef     & BB \nl
NTB 930813.76 & 9212:65850.5 &  76.9 &  77.1 &   8.1 &   3.7 & 336$^{8}$ & abcdefg   & F \nl
NTB 930816.67 & 9215:58569.4 & 155.1 &  53.8 &   5.9 &   3.5 & 429$^{8}$ & bceg      & F \nl
NTB 930820.76 & 9219:65885.3 &  62.7 &  36.4 &   7.6 &   2.8 & 386$^{8}$ & bceg      & F \nl
NTB 930821.64 & 9220:56096.9 & 148.0 & -40.4 &  11.7 &   1.2 & 123$^{1}$ & bc        & F \nl
NTB 930825.48 & 9224:41775.3 &  59.7 &  63.1 &  21.4 &   1.5 & 298$^{4}$ & cg        & F \nl
NTB 930827.60 & 9226:51963.0 & 349.4 &  68.5 &  19.0 &   2.5 & 506$^{8}$ & bceg      & F \nl
NTB 930902.45 & 9232:39001.8 & 224.2 &  20.8 &   6.6 &   2.1 & 364$^{8}$ & acg       & F \nl
NTB 930918.46 & 9248:39913.1 & 275.3 & -81.0 &  11.8 &   2.6 & 310$^{8}$ & bceg      & F \nl
NTB 930921.84 & 9251:73133.2 &  75.0 & -35.7 &  25.2 &   1.3 & 292$^{8}$ & bce       & F \nl
NTB 930924.37 & 9254:32251.0 &  98.1 &  -9.6 &  21.5 &   3.3 & 403$^{8}$ & bcde      & F \nl
NTB 930928.93 & 9258:81199.3 & 260.5 & -67.6 &  10.4 &   2.2 & 529$^{8}$ & bcf       & F \nl
NTB 930928.94 & 9258:81393.3 &  79.8 &  45.2 &   7.0 &   1.7 & 329$^{8}$ & f         & F \nl
NTB 931001.06 & 9261: 5859.5 & 207.2 &  12.6 &   1.4 &   5.6 & 582$^{8}$ & abcde     & R \nl
NTB 931001.72 & 9261:62917.3 &   8.5 &  12.5 &   2.2 &   3.4 & 473$^{8}$ & bcefg     & BB \nl
NTB 931007.20 & 9267:17771.7 & 311.3 &  -5.2 &  23.7 &   2.2 & 368$^{8}$ & bg        & F \nl
NTB 931007.33 & 9267:29319.3 & 355.1 &  36.4 &  48.7 &   1.7 & 363$^{8}$ & bg        & F \nl
NTB 931008.05 & 9268: 5024.9 &  26.4 & -65.6 &  30.5 &   1.2 & 190$^{1}$ & bc        & F \nl
NTB 931008.63 & 9268:54916.8 &  42.3 &  42.8 &   3.7 &   4.7 & 492$^{8}$ & ceg       & BB \nl
NTB 931011.96 & 9271:83718.3 & 248.4 &  63.9 &  23.4 &   1.2 & 373$^{4}$ & bc        & F \nl
NTB 931014.08 & 9274: 7552.2 &  65.5 &  72.0 &  55.6 &   1.5 & 148$^{1}$ & bc        & F \nl
NTB 931017.22 & 9277:19221.7 & 185.7 & -67.5 &  18.3 &   1.8 & 448$^{8}$ & b         & F \nl
NTB 931020.10 & 9280: 8697.5 & 285.0 &  16.0 &  14.0 &   1.7 & 447$^{8}$ & c         & F \nl
NTB 931025.93 & 9285:80462.5 & 173.2 &  15.9 &   4.8 &   2.7 & 438$^{8}$ & bceg      & F \nl
NTB 931031.23 & 9291:20519.6 & 173.2 &  63.4 &  17.2 &   3.0 & 163$^{1}$ & bce       & R \nl
NTB 931106.48 & 9297:42228.4 &  34.3 &  69.7 &   3.7 &   3.7 & 406$^{8}$ & bcefg     & BB \nl
NTB 931106.90 & 9297:78310.6 & 185.7 & -38.0 &  16.1 &   2.7 & 279$^{4}$ & bc        & R \nl
NTB 931107.31 & 9298:26896.0 & 106.0 & -27.7 &  12.6 &   2.9 & 578$^{8}$ & abcde     & BB \nl
NTB 931111.71 & 9302:61941.4 & 244.6 &  49.5 &   0.8 &   9.2 & 352$^{8}$ & abcdefg   & R \nl
NTB 931113.04 & 9304: 3669.7 &  50.9 & -40.2 &  10.1 &   2.1 & 383$^{8}$ & bce       & F \nl
NTB 931115.77 & 9306:66557.6 & 279.1 &  27.9 &  11.0 &   3.1 & 401$^{4}$ & abceg     & R \nl
NTB 931125.86 & 9316:74847.4 & 177.7 & -77.4 &   2.4 &  12.0 & 529$^{8}$ & abcdefg   & R \nl
NTB 931206.45 & 9327:39631.5 & 174.1 & -14.7 &  15.0 &   3.1 & 404$^{8}$ & bceg      & F \nl
NTB 931209.89 & 9330:77266.1 &  20.0 & -37.4 &   3.0 &   3.3 & 443$^{8}$ & bceg      & F \nl
NTB 931215.12 & 9336:10491.5 & 326.0 &  -3.5 &  12.2 &   2.3 & 298$^{4}$ & abd       & F \nl
NTB 931220.16 & 9341:13826.7 & 120.2 &  45.7 &   8.5 &   3.0 & 414$^{8}$ & bce       & F \nl
NTB 931220.73 & 9341:63345.3 &  50.0 &  22.0 &   3.9 &   7.3 & 435$^{8}$ & abcdefg   & R \nl
NTB 931222.11 & 9343:10361.5 & 278.9 &  29.5 &   5.4 &   1.9 & 565$^{8}$ & g         & F \nl
NTB 931222.82 & 9343:70972.1 & 209.6 & -35.5 &  14.8 &   6.2 & 178$^{1}$ & abceg     & R \nl
NTB 931223.07 & 9344: 6589.6 & 212.8 &  40.0 &  23.2 &   1.5 & 200$^{1}$ & bc        & F \nl
\enddata
\end{deluxetable}

\clearpage

\begin{deluxetable}{rrrrr}
\scriptsize
\tablehead{\colhead{Name} &
  \colhead{\begin{tabular}{cc}$T_{50}$\\(s)\end{tabular}} & \colhead{\begin{tabular}{cc}$T_{90}$\\(s)\end{tabular}} &
  \colhead{\begin{tabular}{cc}Peak Flux\\(ph cm$^{-2}$
      s$^{-1}$)\end{tabular}}
  & \colhead{\begin{tabular}{cc}Fluence\\(ergs
      cm$^{-2}$)\end{tabular}}}
\tablecaption{Durations and intensities of non-triggered GRB
  candidates.
\label{tbl:grbs2}}
\startdata
NTB 930118.74 &  2.05 $\pm$  0.00 &  4.10 $\pm$  0.00 &  4.53 $\pm$  0.14 & (3.00 $\pm$ 0.19)E-6 \nl
NTB 930211.88 & 19.46 $\pm$  2.90 & 58.37 $\pm$  6.23 &  0.26 $\pm$  0.05 & (8.46 $\pm$ 1.95)E-7 \nl
NTB 930216.63 & 23.55 $\pm$  0.00 & 40.96 $\pm$  8.44 &  0.28 $\pm$  0.07 & (6.34 $\pm$ 1.54)E-7 \nl
NTB 930217.80 & \nodata &  3.07 $\pm$  1.02 &  0.18 $\pm$  0.10 & (5.86 $\pm$ 3.50)E-8 \nl
NTB 930225.86 & \nodata &  4.10 $\pm$  1.02 &  0.18 $\pm$  0.07 & (7.72 $\pm$ 2.94)E-8 \nl
NTB 930227.83 & 37.89 $\pm$  2.29 & 57.34 $\pm$  7.24 &  0.17 $\pm$  0.05 & (5.46 $\pm$ 1.36)E-7 \nl
NTB 930228.85 & 26.62 $\pm$  3.24 & 91.14 $\pm$ 16.67 &  0.16 $\pm$  0.04 & (9.25 $\pm$ 1.98)E-7 \nl
NTB 930302.20 & 21.50 $\pm$  3.24 & 57.34 $\pm$ 15.50 &  0.26 $\pm$  0.09 & (1.44 $\pm$ 0.44)E-6 \nl
NTB 930303.65 & 32.77 $\pm$  1.45 & 70.66 $\pm$  7.17 &  0.20 $\pm$  0.05 & (7.64 $\pm$ 1.42)E-7 \nl
NTB 930305.70 &  5.12 $\pm$  1.45 & 12.29 $\pm$  1.45 &  0.59 $\pm$  0.19 & (8.88 $\pm$ 3.73)E-7 \nl
NTB 930307.54 &  5.12 $\pm$  1.02 & 22.53 $\pm$ 16.79 &  0.18 $\pm$  0.19 & (2.26 $\pm$ 2.35)E-7 \nl
NTB 930308.30 & 18.43 $\pm$  1.45 & 38.91 $\pm$  2.29 &  0.43 $\pm$  0.11 & (2.18 $\pm$ 0.53)E-6 \nl
NTB 930310.08 & 10.24 $\pm$  0.00 & 34.82 $\pm$ 23.57 &  0.35 $\pm$  0.05 & (4.16 $\pm$ 0.68)E-7 \nl
NTB 930315.46 & 15.36 $\pm$  2.29 & 35.84 $\pm$  4.22 &  0.23 $\pm$  0.05 & (7.61 $\pm$ 1.81)E-7 \nl
NTB 930316.74 & \nodata &  3.07 $\pm$  1.02 &  0.22 $\pm$  0.07 & (8.05 $\pm$ 2.47)E-8 \nl
NTB 930318.18 & 15.36 $\pm$  2.05 & 35.84 $\pm$  6.87 &  0.17 $\pm$  0.04 & (3.97 $\pm$ 0.94)E-7 \nl
NTB 930320.94 & 25.60 $\pm$  0.00 & 59.39 $\pm$  3.07 &  0.96 $\pm$  0.15 & (4.09 $\pm$ 0.62)E-6 \nl
NTB 930325.65 & 10.24 $\pm$  1.45 & 20.48 $\pm$  5.12 &  0.24 $\pm$  0.06 & (5.13 $\pm$ 1.42)E-7 \nl
NTB 930327.46 & 12.29 $\pm$  3.69 & 46.08 $\pm$ 31.89 &  0.29 $\pm$  0.07 & (7.86 $\pm$ 2.88)E-7 \nl
NTB 930330.91 & 12.29 $\pm$  2.29 & 52.22 $\pm$ 14.37 &  0.24 $\pm$  0.06 & (4.99 $\pm$ 1.47)E-7 \nl
NTB 930403.84 & 13.31 $\pm$  0.00 & 92.16 $\pm$ 10.69 &  0.26 $\pm$  0.04 & (6.96 $\pm$ 0.93)E-7 \nl
NTB 930409.13 & 14.34 $\pm$  1.02 & 39.94 $\pm$  9.27 &  0.40 $\pm$  0.05 & (7.91 $\pm$ 1.20)E-7 \nl
NTB 930409.91 & 12.29 $\pm$  1.02 & 53.25 $\pm$  0.00 &  1.19 $\pm$  0.06 & (3.06 $\pm$ 0.34)E-6 \nl
NTB 930410.76 & 24.58 $\pm$  1.02 & 74.75 $\pm$  3.07 &  0.29 $\pm$  0.05 & (1.09 $\pm$ 0.16)E-6 \nl
NTB 930416.56 & \nodata & 20.48 $\pm$  5.12 &  0.21 $\pm$  0.04 & (3.54 $\pm$ 0.58)E-7 \nl
NTB 930417.78 &  5.12 $\pm$  1.45 & 35.84 $\pm$ 26.62 &  0.21 $\pm$  0.06 & (2.05 $\pm$ 0.82)E-7 \nl
NTB 930421.11 & 47.10 $\pm$  3.24 & (1.29 $\pm$ 0.23)E+2 &  0.15 $\pm$  0.04 & (1.28 $\pm$ 0.26)E-6 \nl
NTB 930422.58 & 15.36 $\pm$  1.02 & 44.03 $\pm$ 17.41 &  0.27 $\pm$  0.03 & (1.20 $\pm$ 0.13)E-6 \nl
NTB 930424.45 & 41.98 $\pm$  1.02 & (1.06 $\pm$ 0.03)E+2 &  1.28 $\pm$  0.06 & (1.38 $\pm$ 0.08)E-5 \nl
NTB 930424.97 & 12.29 $\pm$  1.02 & 32.77 $\pm$ 15.36 &  0.33 $\pm$  0.04 & (8.77 $\pm$ 1.30)E-7 \nl
NTB 930426.48 & \nodata &  3.07 $\pm$  1.02 &  0.70 $\pm$  0.10 & (1.66 $\pm$ 0.27)E-7 \nl
NTB 930427.59 & 21.50 $\pm$  1.45 & 46.08 $\pm$  7.17 &  0.19 $\pm$  0.06 & (4.45 $\pm$ 1.30)E-7 \nl
NTB 930429.75 & 43.01 $\pm$  1.45 & (1.11 $\pm$ 0.15)E+2 &  0.38 $\pm$  0.04 & (2.24 $\pm$ 0.25)E-6 \nl
NTB 930501.34 & \nodata &  2.05 $\pm$  1.02 &  0.23 $\pm$  0.14 & (5.71 $\pm$ 3.59)E-8 \nl
NTB 930506.63 & 34.82 $\pm$  8.26 & 82.94 $\pm$  5.22 &  0.18 $\pm$  0.05 & (9.14 $\pm$ 2.73)E-7 \nl
NTB 930508.95 &  5.12 $\pm$  1.45 & 22.53 $\pm$ 15.39 &  0.24 $\pm$  0.05 & (4.17 $\pm$ 1.39)E-7 \nl
NTB 930513.98 & \nodata & 41.98 $\pm$ 10.24 &  0.17 $\pm$  0.05 & (3.67 $\pm$ 1.07)E-7 \nl
NTB 930519.39 & \nodata &  1.02 $\pm$  0.00 &  0.35 $\pm$  0.29 & (1.15 $\pm$ 0.96)E-7 \nl
NTB 930612.63 &  6.14 $\pm$  2.05 & 50.18 $\pm$  9.22 &  1.33 $\pm$  0.08 & (1.23 $\pm$ 0.42)E-6 \nl
NTB 930616.27 & \nodata &  1.02 $\pm$  0.00 &  3.21 $\pm$  0.77 & (7.97 $\pm$ 1.94)E-7 \nl
NTB 930617.23 & 51.20 $\pm$  1.02 & 73.73 $\pm$  5.12 &  0.29 $\pm$  0.06 & (1.35 $\pm$ 0.28)E-6 \nl
NTB 930626.94 & 14.34 $\pm$  1.02 & 61.44 $\pm$ 38.01 &  0.27 $\pm$  0.05 & (8.29 $\pm$ 1.25)E-7 \nl
NTB 930630.71 & \nodata & 18.43 $\pm$  5.12 &  0.23 $\pm$  0.05 & (3.86 $\pm$ 0.71)E-7 \nl
NTB 930701.62 & 20.48 $\pm$  1.02 & 35.84 $\pm$  3.07 &  0.45 $\pm$  0.05 & (1.07 $\pm$ 0.14)E-6 \nl
NTB 930705.64 & 12.29 $\pm$  1.02 & 48.13 $\pm$ 16.38 &  0.20 $\pm$  0.05 & (4.10 $\pm$ 0.95)E-7 \nl
NTB 930717.20 &  2.05 $\pm$  1.45 &  3.07 $\pm$  1.02 &  0.21 $\pm$  0.09 & (1.37 $\pm$ 1.11)E-7 \nl
NTB 930717.98 &  6.14 $\pm$  1.45 & 17.41 $\pm$  3.07 &  0.32 $\pm$  0.05 & (3.93 $\pm$ 1.09)E-7 \nl
NTB 930722.84 & 26.62 $\pm$  2.29 & 72.70 $\pm$  9.44 &  0.20 $\pm$  0.05 & (8.37 $\pm$ 1.62)E-7 \nl
NTB 930728.54 & 66.56 $\pm$ 15.22 & (3.43 $\pm$  0.21)E+2 &  0.29 $\pm$  0.05 & (3.60 $\pm$ 0.92)E-6 \nl
NTB 930804.71 & 19.46 $\pm$  1.02 & 49.15 $\pm$ 16.42 &  0.25 $\pm$  0.04 & (9.05 $\pm$ 1.10)E-7 \nl
NTB 930811.62 & 18.43 $\pm$  2.29 & 45.06 $\pm$  4.22 &  0.29 $\pm$  0.06 & (7.27 $\pm$ 1.71)E-7 \nl
NTB 930812.27 & 28.67 $\pm$  2.29 & 77.82 $\pm$  3.24 &  0.39 $\pm$  0.04 & (1.97 $\pm$ 0.25)E-6 \nl
NTB 930813.76 & (1.02 $\pm$ 0.02)E+2 & (1.36 $\pm$ 0.02)E+2 &  0.29 $\pm$  0.05 & (1.16 $\pm$ 0.19)E-6 \nl
NTB 930816.67 &  8.19 $\pm$  1.02 & 24.58 $\pm$  4.10 &  0.33 $\pm$  0.10 & (6.95 $\pm$ 2.21)E-7 \nl
NTB 930820.76 &  7.17 $\pm$  1.02 & 45.06 $\pm$ 23.55 &  0.22 $\pm$  0.05 & (4.03 $\pm$ 0.98)E-7 \nl
NTB 930821.64 & \nodata &  2.05 $\pm$  1.02 &  0.30 $\pm$  0.10 & (9.30 $\pm$ 3.26)E-8 \nl
NTB 930825.48 &  4.10 $\pm$  2.29 & 20.48 $\pm$  5.12 &  0.14 $\pm$  0.06 & (1.58 $\pm$ 1.07)E-7 \nl
NTB 930827.60 & 15.36 $\pm$  3.07 & 48.13 $\pm$ 15.36 &  0.22 $\pm$  0.07 & (5.23 $\pm$ 1.80)E-7 \nl
NTB 930902.45 & 28.67 $\pm$ 10.69 & 74.75 $\pm$ 16.51 &  0.16 $\pm$  0.04 & (7.53 $\pm$ 3.05)E-7 \nl
NTB 930918.46 & 16.38 $\pm$  4.10 & 54.27 $\pm$ 14.34 &  0.20 $\pm$  0.05 & (5.54 $\pm$ 1.95)E-7 \nl
NTB 930921.84 & 12.29 $\pm$  3.24 & 56.32 $\pm$ 18.43 &  0.32 $\pm$  0.12 & (3.12 $\pm$ 1.47)E-7 \nl
NTB 930924.37 &  8.19 $\pm$  1.02 & 28.67 $\pm$ 17.17 &  0.22 $\pm$  0.05 & (5.58 $\pm$ 1.23)E-7 \nl
NTB 930928.93 &  2.05 $\pm$  1.02 &  4.10 $\pm$  2.05 &  0.19 $\pm$  0.06 & (6.92 $\pm$ 4.24)E-8 \nl
NTB 930928.94 & 14.34 $\pm$  1.45 & 28.67 $\pm$  2.29 &  0.13 $\pm$  0.04 & (3.75 $\pm$ 1.04)E-7 \nl
NTB 931001.06 &  4.10 $\pm$  0.00 & 12.29 $\pm$  3.07 &  0.74 $\pm$  0.08 & (8.66 $\pm$ 0.93)E-7 \nl
NTB 931001.72 & 16.38 $\pm$  2.29 & 82.94 $\pm$  8.44 &  0.28 $\pm$  0.06 & (9.18 $\pm$ 1.98)E-7 \nl
NTB 931007.20 &  5.12 $\pm$  1.02 & 35.84 $\pm$  5.12 &  0.23 $\pm$  0.07 & (3.62 $\pm$ 1.27)E-7 \nl
NTB 931007.33 & \nodata &  8.19 $\pm$  4.10 &  0.43 $\pm$  0.74 & (4.94 $\pm$ 8.59)E-7 \nl
NTB 931008.05 & \nodata &  2.05 $\pm$  1.02 &  0.28 $\pm$  0.11 & (1.19 $\pm$ 0.46)E-7 \nl
NTB 931008.63 & 66.56 $\pm$  1.45 & (1.98 $\pm$ 0.28)E+2 &  0.26 $\pm$  0.04 & (3.34 $\pm$ 0.29)E-6 \nl
NTB 931011.96 & 11.26 $\pm$  3.24 & 50.18 $\pm$ 18.55 &  0.20 $\pm$  0.11 & (2.00 $\pm$ 1.27)E-7 \nl
NTB 931014.08 & \nodata &  4.10 $\pm$  2.05 &  0.36 $\pm$  0.14 & (1.24 $\pm$ 0.52)E-7 \nl
NTB 931017.22 &  9.22 $\pm$  5.22 & 41.98 $\pm$ 23.57 &  0.16 $\pm$  0.06 & (2.72 $\pm$ 1.82)E-7 \nl
NTB 931020.10 & 21.50 $\pm$  3.24 & 73.73 $\pm$  2.29 &  0.14 $\pm$  0.05 & (4.52 $\pm$ 1.52)E-7 \nl
NTB 931025.93 & 20.48 $\pm$  9.22 & 49.15 $\pm$ 15.66 &  0.21 $\pm$  0.05 & (5.15 $\pm$ 2.55)E-7 \nl
NTB 931031.23 & \nodata &  1.02 $\pm$  0.00 &  0.40 $\pm$  0.11 & (8.95 $\pm$ 2.43)E-8 \nl
NTB 931106.48 & 31.74 $\pm$  2.29 & 90.11 $\pm$  1.45 &  0.22 $\pm$  0.04 & (1.41 $\pm$ 0.21)E-6 \nl
NTB 931106.90 & \nodata & 11.26 $\pm$  4.10 &  0.32 $\pm$  0.06 & (2.66 $\pm$ 0.50)E-7 \nl
NTB 931107.31 & 30.72 $\pm$  3.24 & 55.30 $\pm$  9.27 &  0.41 $\pm$  0.11 & (1.57 $\pm$ 0.40)E-6 \nl
NTB 931111.71 & 58.37 $\pm$  3.07 & (1.86 $\pm$ 0.01)E+2 &  0.77 $\pm$  0.04 & (7.24 $\pm$ 0.59)E-6 \nl
NTB 931113.04 & 12.29 $\pm$  1.02 & 52.22 $\pm$ 16.67 &  0.21 $\pm$  0.05 & (4.15 $\pm$ 0.93)E-7 \nl
NTB 931115.77 & 14.34 $\pm$  1.45 & 26.62 $\pm$  2.29 &  0.32 $\pm$  0.06 & (3.15 $\pm$ 0.76)E-7 \nl
NTB 931125.86 &  5.12 $\pm$  1.02 & 14.34 $\pm$  1.45 &  0.83 $\pm$  0.05 & (1.27 $\pm$ 0.27)E-6 \nl
NTB 931206.45 & 11.26 $\pm$  1.02 & 29.70 $\pm$  8.44 &  0.26 $\pm$  0.05 & (5.43 $\pm$ 1.11)E-7 \nl
NTB 931209.89 & 34.82 $\pm$  1.02 & 90.11 $\pm$  4.58 &  0.27 $\pm$  0.06 & (1.44 $\pm$ 0.21)E-6 \nl
NTB 931215.12 &  3.07 $\pm$  2.05 & 13.31 $\pm$ 11.31 &  0.28 $\pm$  0.07 & (2.25 $\pm$ 1.61)E-7 \nl
NTB 931220.16 &  4.10 $\pm$  1.45 & 11.26 $\pm$  4.10 &  0.49 $\pm$  0.11 & (6.84 $\pm$ 2.81)E-7 \nl
NTB 931220.73 &  8.19 $\pm$  0.00 & 33.79 $\pm$  4.58 &  0.56 $\pm$  0.05 & (9.78 $\pm$ 0.89)E-7 \nl
NTB 931222.11 & 18.43 $\pm$  3.24 & 58.37 $\pm$ 12.83 &  0.18 $\pm$  0.06 & (6.86 $\pm$ 1.93)E-7 \nl
NTB 931222.82 & \nodata &  2.05 $\pm$  1.02 &  1.75 $\pm$  0.24 & (4.40 $\pm$ 0.65)E-7 \nl
NTB 931223.07 & \nodata &  3.07 $\pm$  1.02 &  0.26 $\pm$  0.19 & (9.55 $\pm$ 7.28)E-8 \nl
\enddata
\end{deluxetable}

\clearpage
\setcounter{figure}{0}
\begin{figure}
\plotone{f1.eps}
\caption{~}
\end{figure}

\begin{figure}
\plotone{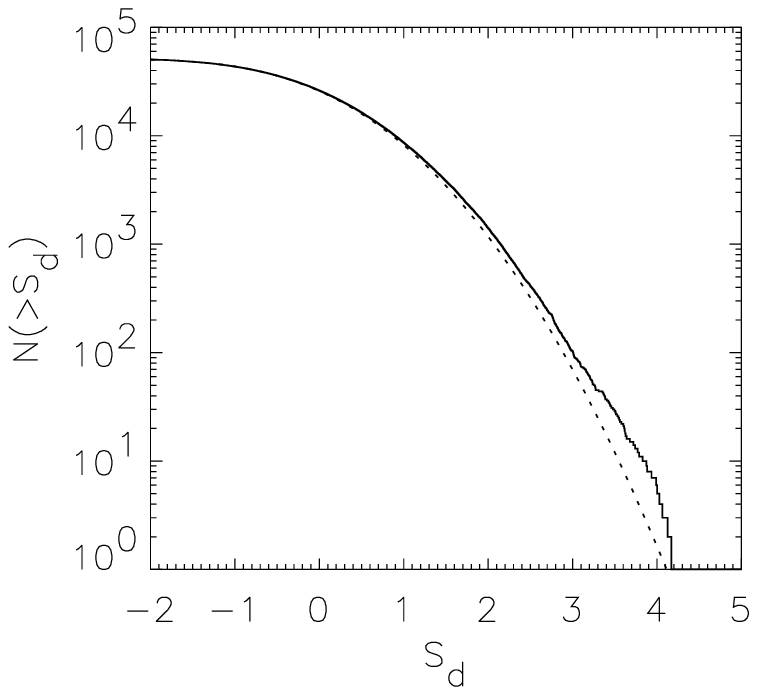}
\caption{~}
\end{figure}

\begin{figure}
\plotone{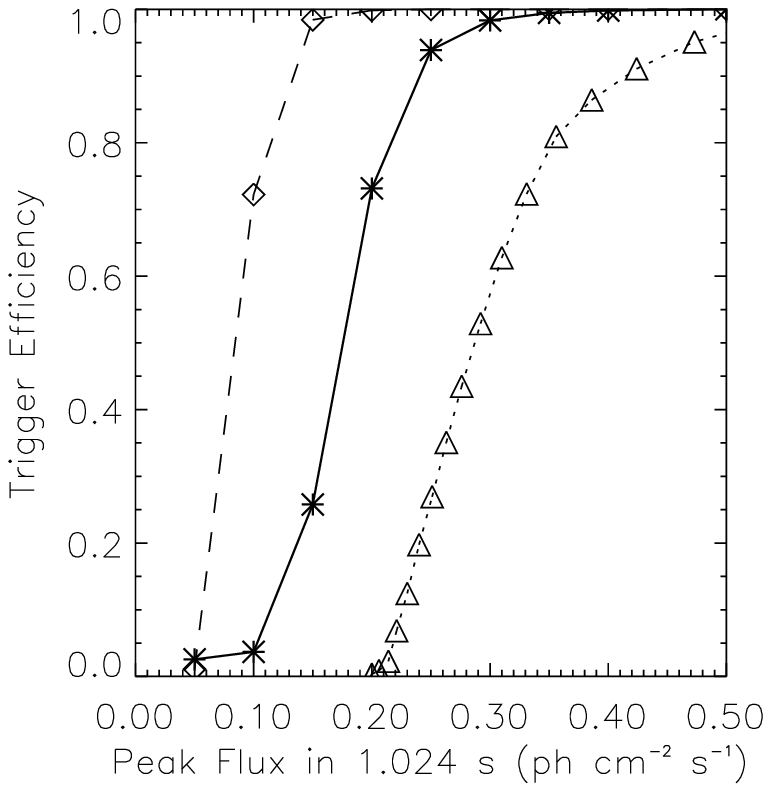}
\caption{~}
\end{figure}

\begin{figure}
\plotone{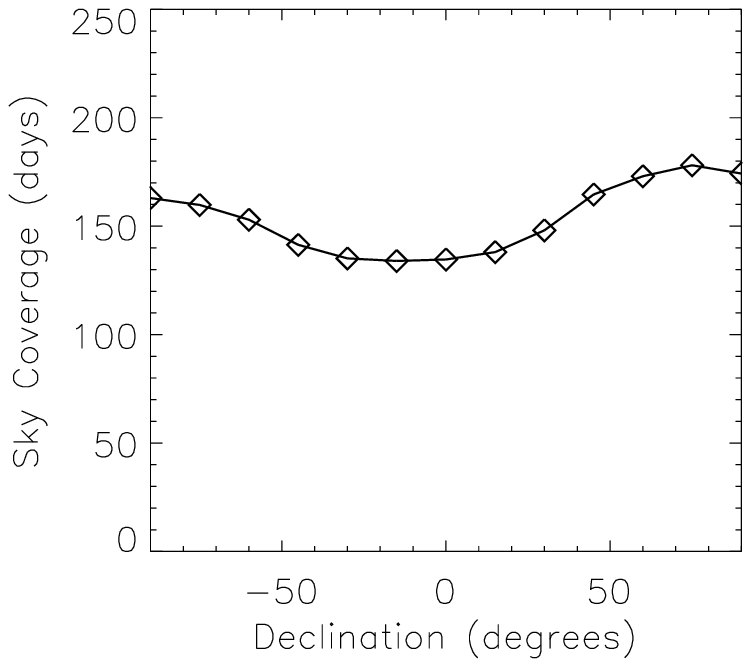}
\caption{~}
\end{figure}

\begin{figure}
\plotone{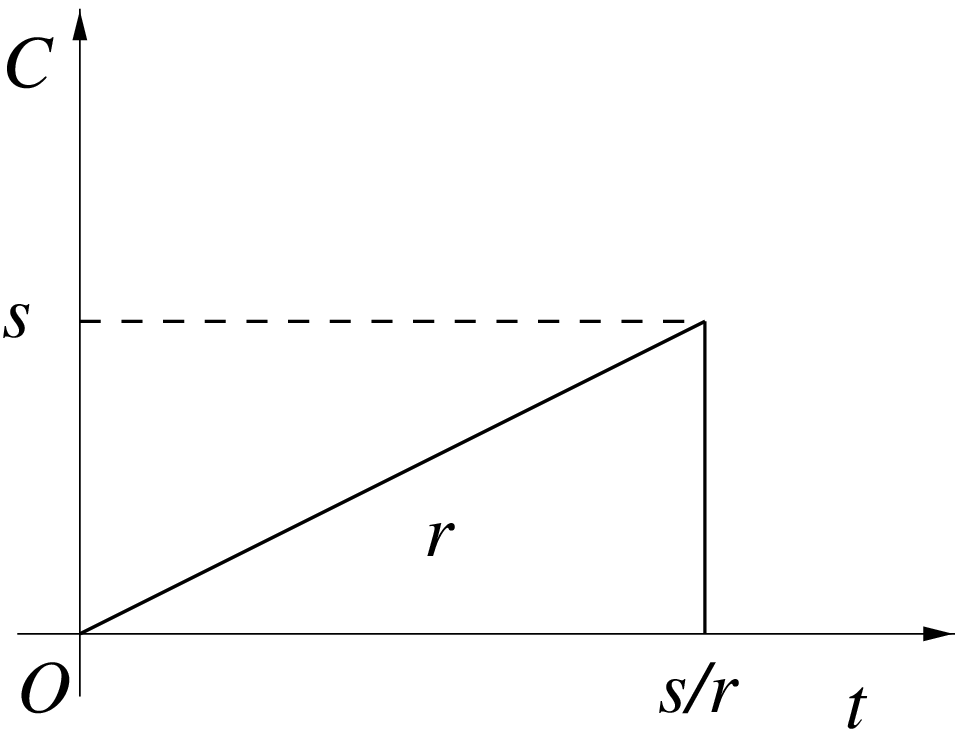}
\caption{~}
\end{figure}

\begin{figure}
\plotone{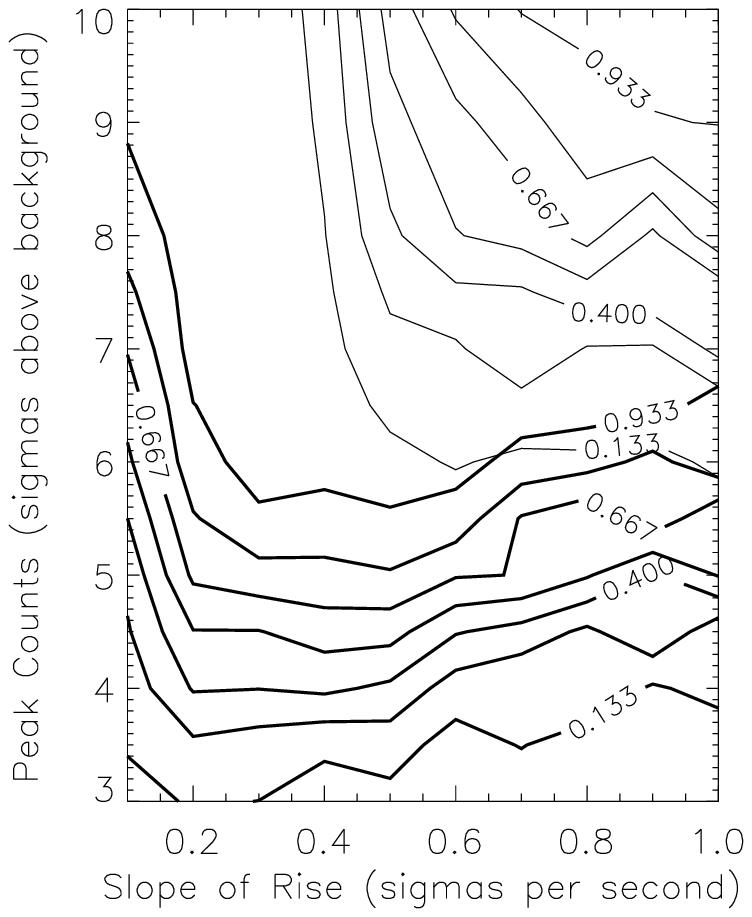}
\caption{~}
\end{figure}
  
\begin{figure}
\plotone{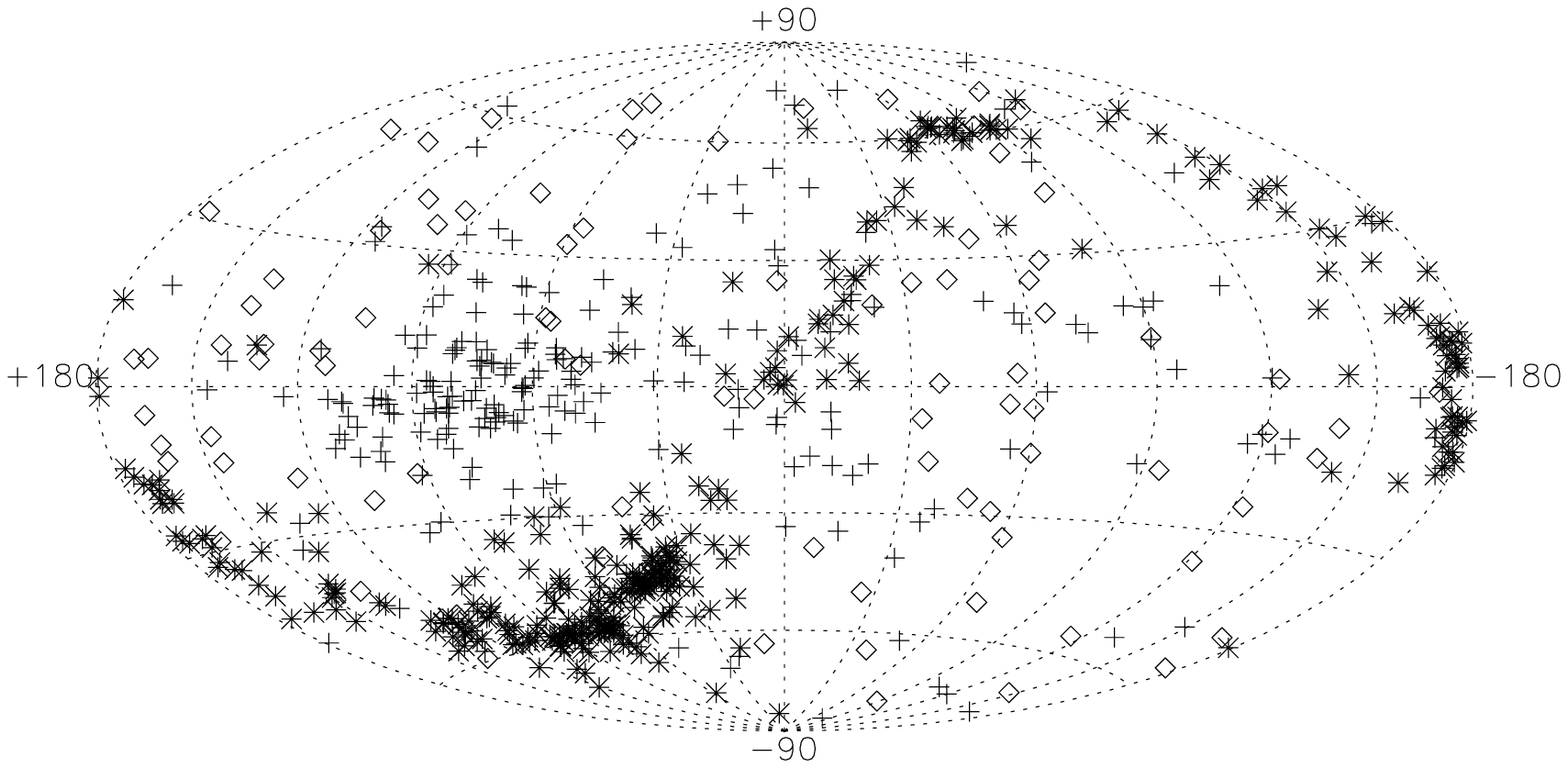}
\caption{~}
\end{figure}

\begin{figure}
\plotone{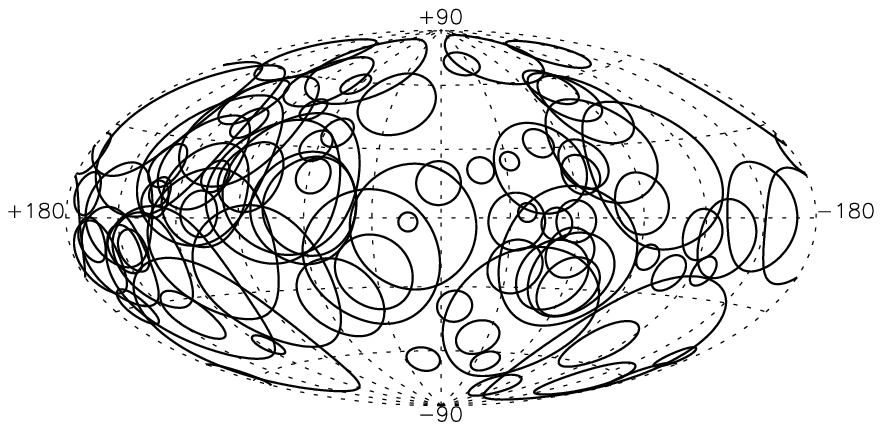}
\caption{~}
\end{figure}

\begin{figure}
\plotone{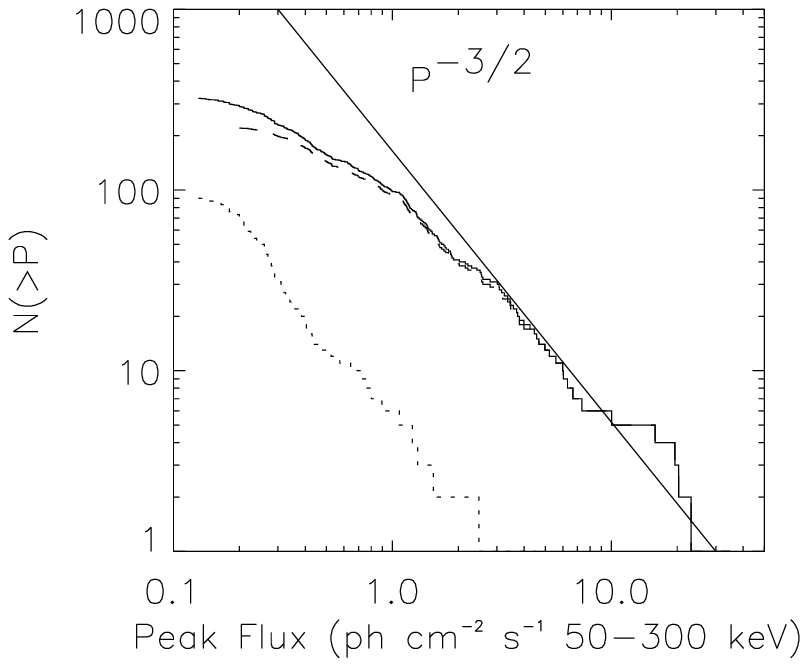}
\caption{~}
\end{figure}

\begin{figure}
\epsscale{0.6}
\plotone{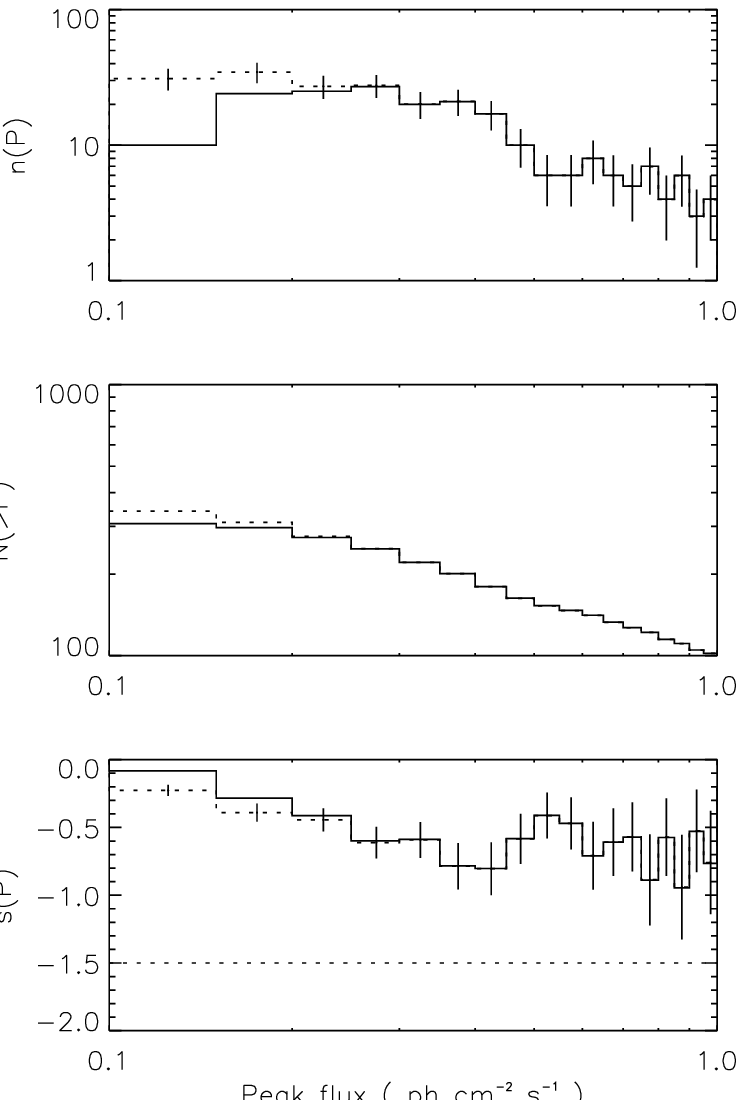}
\caption{~}
\end{figure}

\begin{figure}
\epsscale{1.0}
\plotone{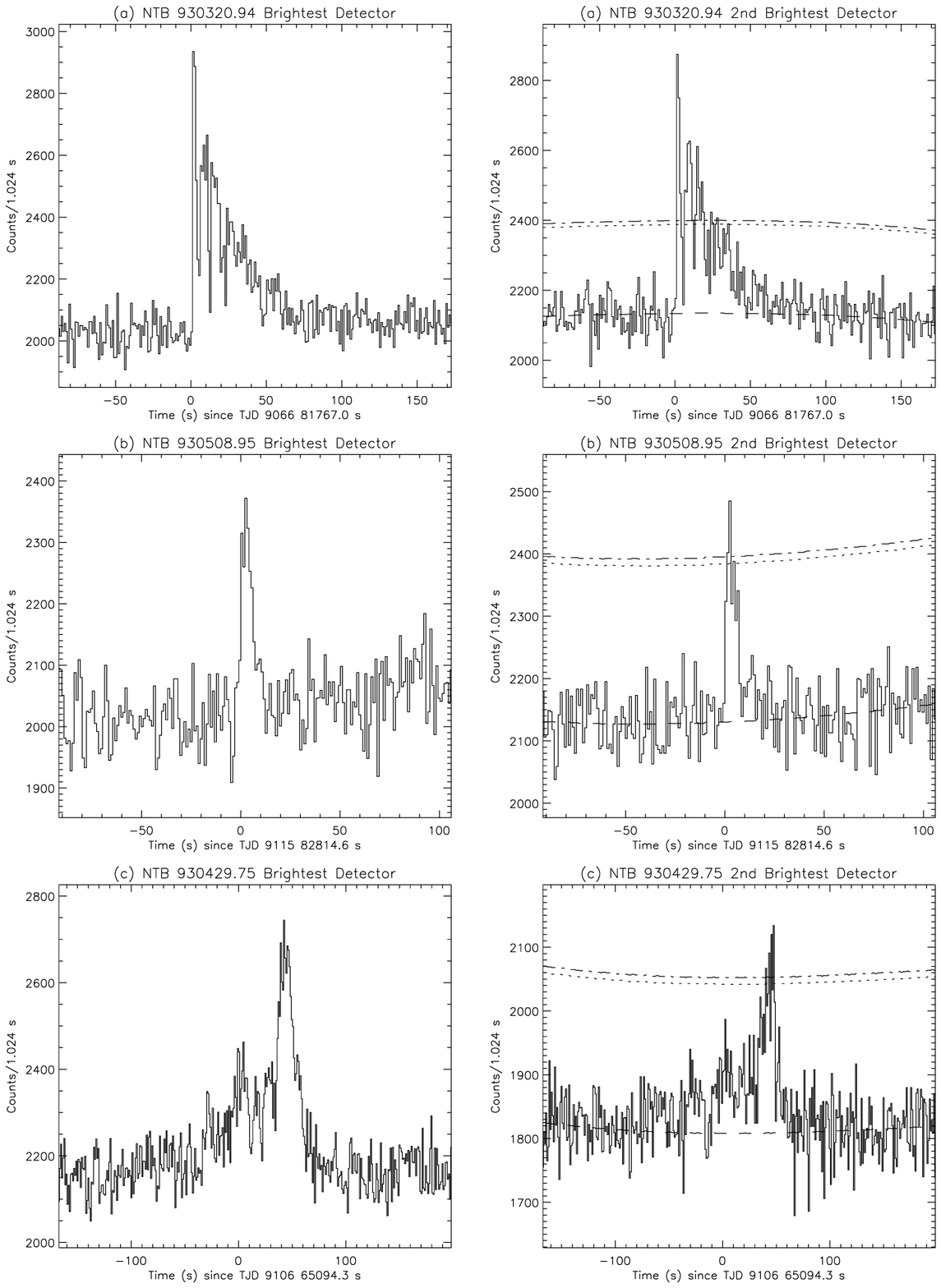}
\caption{page 1}
\end{figure}
\setcounter{figure}{10}
\begin{figure}
\plotone{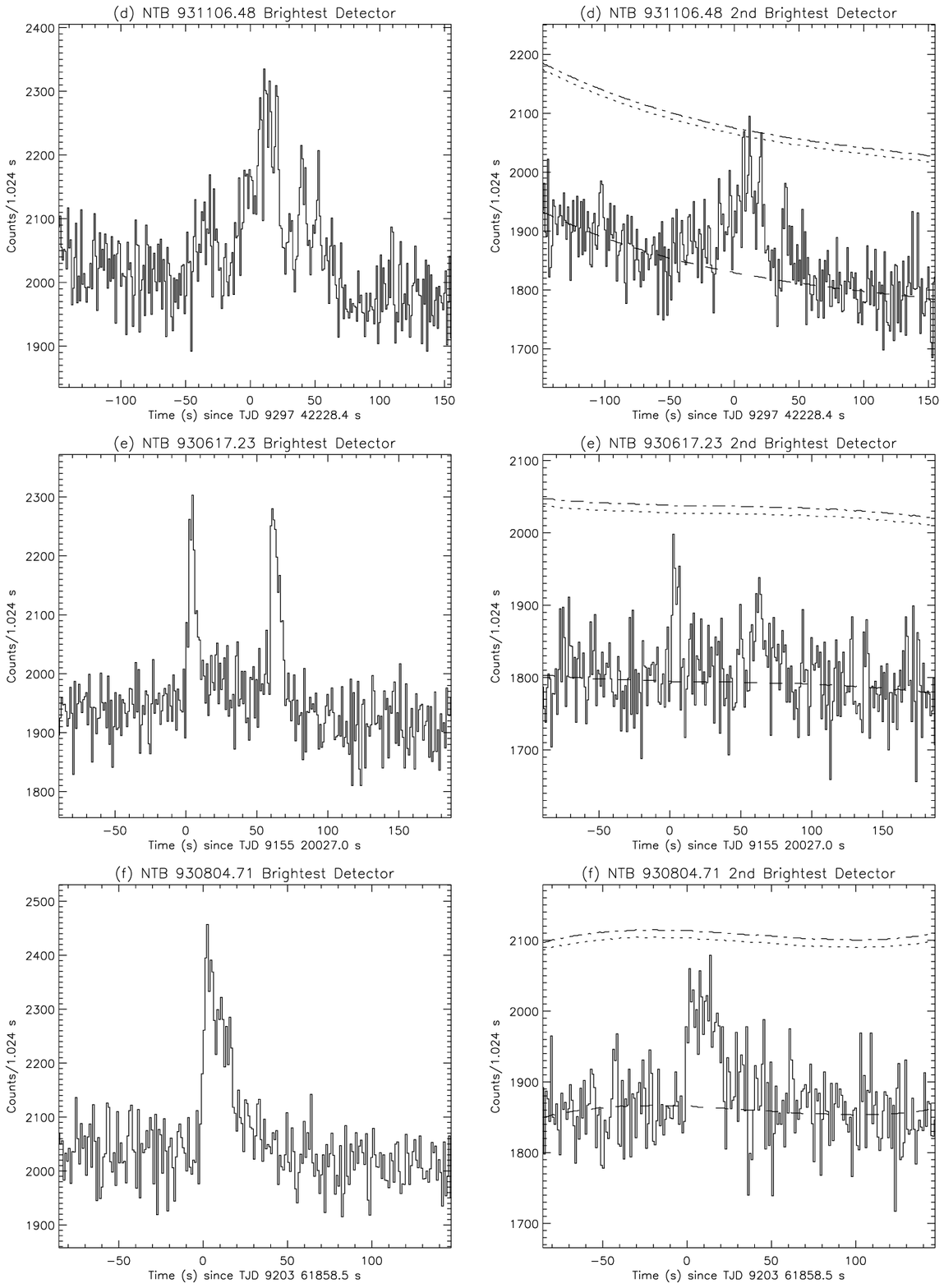}
\caption{page 2}
\end{figure}

\begin{figure}
\epsscale{1.0} 
\plotone{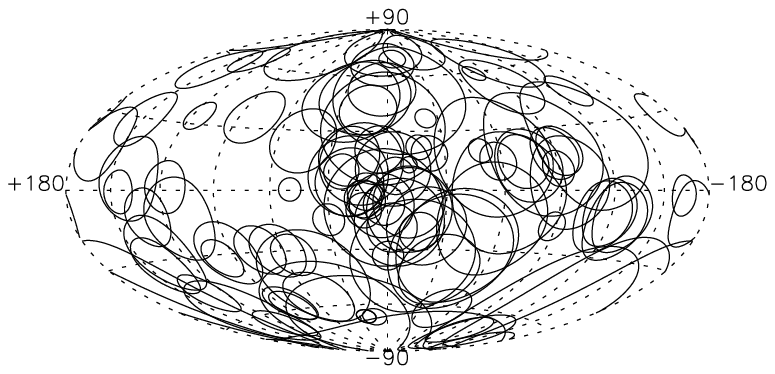}
\caption{~}
\end{figure}

\end{document}